\documentclass[fleqn,usenatbib]{mn2e}
\usepackage{mathptmx}
\usepackage{txfonts}
\usepackage[T1]{fontenc}

\usepackage{graphicx}

\newcommand\pp{{\phantom{$+$}}}

\newcommand\teff{{T_{\rm eff}}}
\newcommand\lta{\mathrel{\hbox{\raise 0.6 ex \hbox{$<$}\kern
                   -1.8 ex\lower .5 ex\hbox{$\sim$}}}}
\newcommand\gta{\mathrel{\hbox{\raise 0.6 ex \hbox{$>$}\kern
                   -1.7 ex\lower .5 ex\hbox{$\sim$}}}}

\title[Isochrones for Several Mixtures of the Metals]{Models for Metal-Poor
Stars with Different Initial Abundances of C, N, O, Mg, and Si.~II.~Application
to the Colour-Magnitude Diagrams of the Globular Clusters 47 Tuc, NGC\,6362,
M\,5, M\,3, M\,55, and M$\,$92}

\author[D. A. VandenBerg et al.]{
Don A.~VandenBerg,$^{1}$\thanks{E-mail: vandenbe@uvic.ca } 
Luca Casagrande,$^{2}$
and Bengt Edvardsson,$^{3}$ 
\\
$^{1}$Department of Physics and Astronomy, University of Victoria, 
P.O. Box 1700, STN CSC, Victoria, BC, Canada V8W 2Y2 \\
$^{2}$Research School of Astronomy and Astrophysics, Mount Stromlo Observatory,
The Australian National University, Canberra, A.C.T.~2611, Australia \\
$^{3}$Theoretical Astrophysics, Department of Physics \& Astronomy,
Uppsala University, Box 516, SE-751 20 Uppsala, Sweden }

\date{Accepted XXX. Received YYY; in original form ZZZ}
\pubyear{2020}

\begin{document}
\label{firstpage}
\pagerange{\pageref{firstpage}--\pageref{lastpage}}
\maketitle

\begin{abstract}

Stellar models for $-2.5 \le$ [Fe/H] $\le -0.5$ that have been computed for
variations in the C:N:O abundance ratio (for two different values of [CNO/Fe])
are compared with {\it HST} Wide Field Camera 3 (WFC3) observations of the
globular clusters (GCs) 47 Tuc, NGC\,6362, M\,5 M\,3, M\,55, and M\,92.  The BCs
used to transpose the models to the observed planes are based on new MARCS
synthetic spectra that incorporate improved treatments of molecules that involve
atoms of C, N, and O.  On the assumption of well-supported distance moduli and
reddenings, isochrones for [O/Fe] $= 0.6$ and [$m$/Fe] $= 0.4$ for the other
$\alpha$ elements, which are favoured by binary stars in GCs, generally
reproduce the main features of observed colour-magnitude diagrams (CMDs) to
within $\sim 0.03$ mag.  In particular, they appear to match the spreads in
the observed $(M_{F336W}-M_{F438W})_0$ colours that are spanned by CN-weak and
CN-strong stars along the lower giant branch quite well, but not the bluest
giants, which are suspected to be N-poor ([N/Fe] $\lta -0.5$).  Both the
absolute $(M_{F438W}-M_{F606W})_0$ colours and the variations in these colours
at a given $M_{F606W}$ magnitude on the giant branch are difficult to explain
unless the reddest stars are C-rich ([C/Fe] $\gta +0.5$).  Allowing for moderate
He abundance variations ($\delta\,Y \sim 0.05$) improves the fits to the
observations.

\end{abstract}


\section{Introduction}
\label{sec:intro}
 
Globular clusters (GCs) are known to contain multiple, chemically distinct,
stellar populations.  This has been established by the vast amount of
spectroscopic work that has been carried out during the past 50$+$ years
(see, e.g., the reviews by \citealt{fn81}, \citealt{smi87}, \citealt{kra94},
\citealt{gsc04}), and by numerous photometric surveys since 2004 (e.g.,
\citealt{bpa04}, \citealt{pba07}, \citealt{mpk10}, \citealt{bbp10}).  Chemical
abundance studies have shown that O--Na and Mg--Al anticorrelations are
typically found in GCs (\citealt[and references therein]{cbg09b}), along with
the ubiquitous variations of CN (e.g., \citealt{sn93}, \citealt{gva98},
\citealt{bhs04}, \citealt{cbs05}).  Statistically significant Al--Si
correlations have also been detected in a few clusters (\citealt{ygn05},
\citealt{cbg09b}), as well as star-to-star variations of the isotopic ratios of
Mg (\citealt{she96}, \citealt{yal06}).  All of these findings can be explained
by $p$-capture processes, provided that the temperatures in the nucleosynthesis
site can reach sufficiently high values ($\gta 7 \times 10^7$ K in order for 
the Mg-Al cycle to operate). 

Photometry has the important advantages over spectroscopy in giving one the
capability to obtain observations of tens of thousands of stars simultaneously
and to reach very faint luminosities.  In recent years, multi-wavelength
investigations have produced stunning colour-magnitude diagrams (CMDs) for many
GCs, consisting of separate sequences of stars that can sometimes be
distinguished all of the way from near the bottom of the main sequence (MS) to
the upper red-giant branch (RGB); see, e.g.,
\citet[\citealt{mmp13},\citealt{mmp15a},\citealt{mmp15b}]{mpb12}, \citet{bma17}.
Moreover, the so-called ``chromosome maps" (\citealt{mpr17}) provide the means
to clearly distinguish between different stellar populations, as well as 
sub-populations, in many GCs.  For the most part, these separations are caused
by chemical differences (mainly of C and N), though, with the right filter
choices, stellar populations that have different abundances of Mg can also be
revealed (\citealt{mmr20}).  Importantly, precise photometry can yield tight
constraints on He abundance variations within GCs (\citealt{mmp12},
\citealt{kbc12}, \citealt{npm15}), which are otherwise very difficult to
determine.

\begin{table*}
\centering
\caption{The Adopted Chemical Abundance Mixtures} 
 \label{tab:t1}
\smallskip
\begin{tabular}{lcccccccc}
\hline
\hline
\noalign{\smallskip}
 Names of & & & & & & & & \\
 BC Tables$^{a}$ & He & C & N & O & [CNO/Fe] & Mg & Si & $v_T^{b}$ \\
\noalign{\smallskip}
\hline
\noalign{\smallskip}
 \bf{a4s08}  & 10.93 & \pp8.39 & \pp7.78 & \pp9.06 & $+0.28$ &
   \pp7.93 & \pp7.91 & 2.0 \\
 a4s21  & 11.00 & \pp8.43 & \pp7.83 & \pp9.09 & $+0.28$ &
   \pp8.00 & \pp7.91 & f(g) \\
 a4CN   & \bf{--} & $\downarrow\,0.3$ & $\uparrow\,0.50$ & \bf{--} &
  $+0.28$ & \bf{--} & \bf{--} & f(g) \\
 a4CNN  & \bf{--} & $\downarrow\,0.3$ & $\uparrow\,1.13$ & \bf{--} & $+0.44$ &
   \bf{--} & \bf{--} & f(g) \\
 a4ON   & \bf{--} & $\downarrow\,0.8$ & $\uparrow\,1.30$ & $\downarrow\,0.8$ &
   $+0.28$ & \bf{--} & \bf{--} & f(g) \\
 a4ONN   & \bf{--} & $\downarrow\,0.8$ & $\uparrow\,1.48$ & $\downarrow\,0.8$ &
   $+0.44$ & \bf{--} & \bf{--} & f(g) \\
 a4xC$\_$p4   & \bf{--} & $\uparrow\,0.4$ & \bf{--} & \bf{--} & $+0.38$ &
   \bf{--} & \bf{--} & f(g) \\
 a4xCO   & \bf{--} & $\uparrow\,0.7$ & \bf{--} & $\uparrow\,0.2$ & $+0.61$ &
   \bf{--} & \bf{--} & f(g) \\
 a4xO$\_$p2  & \bf{--} & \bf{--} & \bf{--} & $\uparrow\,0.2$ & $+0.44$ &
   \bf{--} & \bf{--} & f(g) \\ 
 a4xO$\_$p4  & \bf{--} & \bf{--} & \bf{--} & $\uparrow\,0.4$ & $+0.62$ &
   \bf{--} & \bf{--} & f(g) \\ 
\noalign{\smallskip}
\hline
\noalign{\smallskip}
\end{tabular}
\begin{minipage}{1\textwidth}
$^{a}$~Boldface font identifies reference models (see the text); the others
involve changes to the abundances of one or more of the metals, as
tabulated, or to $v_T$.  \\ 
$^{b}$~f(g) implies that the microturbulent velocity, $v_T$, varies with gravity
such that $v_T = 1.0$ km/s if $\log\,g \ge 4.0$ or 2.0 km/s if $\log\,g \le
3.0$. \\
\phantom{~~~~~~~~~~~~~~~~~~~~~~~~~}
\end{minipage}
\end{table*}

Helium differs from the metals that are often described as the ``light elements"
(those below the iron-peak elements in the Periodic Table) in that it mainly
affects the temperatures and luminosities of stars.  At a given effective
temperature ($\teff$), the bolometric corrections (BCs) that are relevant to old
stellar populations have very little sensitivity to the He abundance
(\citealt{gcb07}).  On the other hand, stellar evolutionary computations (see,
e.g., \citealt{swf06}, \citealt{pcs09}, \citealt{vbd12}, \citealt{cmp13}) have
shown that the locations of stars on the H-R diagram are essentially independent
of the abundances of the light elements if, as seems to be the case in most GCs
(e.g., \citealt{isk99}, \citealt{cm05}), C$+$N$+$O and Mg$+$Al$+$Si are
approximately constant.   However, BCs can be very dependent on the actual
abundance of each metal (depending on which spectral features are located within
the passbands of the selected filters and the strengths of those features).  For
instance, variations in the abundances of C, N, and O, as manifested through
the formation of CN, CH, NH, OH, and other molecules, can have significant
effects on the fluxes in the Johnson-Cousins $U$ and $B$ filters (see
\citealt[their Fig.~4]{ssw11}).  Thus the complex morphologies and the large
colour spreads that characterize many recent CMDs are almost entirely the
consequence of BCs instead of variations in $\teff$.

The Wide Field Camera 3 (WFC3) photometry that has been obtained via the
{\it HST} UV Legacy Survey (\citealt{pmb15}, \citealt[hereafter NLP18]{nlp18})
provides a tremendous resource for the testing, and application, of stellar
models.  This survey was designed to use the $F275W$, $F336W$, $F438W$, $F606W$,
and $F814W$ filters to obtain observations that would discriminate between stars
with different abundances of He, C, N, and O.  The availability of these data
provided the main motivation for \citet[hereafter Paper I]{vec21} to compute
stellar models for several different mixtures of the metals, including, in 
particular, different C:N:O ratios for two values of [CNO/Fe].  

These computations employ BCs that are based on new MARCS high-resolution
synthetic spectra which incorporate improved treatments of molecules that
involve atoms of C, N, and O.  The first application of these models, to the
metal-rich GC NGC\,6496, indicated that isochrones which employ the new BCs are
able to reproduce observed colours at UV and optical wavelengths very well,
possibly to within the uncertainties associated with the zero points of both
the synthetic and observed photometry and the cluster properties (distance,
reddening, metallicity).  By comparison, isochrones employing the BCs derived
from the previous generation of MARCS spectra (\citealt{gee08}) fail to match
the observed $m_{F336W}-m_{F438W}$ colours by $\gta 0.12$ mag (though much
smaller differences are obtained for redder colours).  In addition, as shown
by \citet{mcc17}, isochrones coupled with BCs based on Kurucz ATLAS12 model
atmospheres (\citealt{ku14}) and synthetic spectra as produced by the SYNTHE
code (\citealt{ku05}) appear to predict too much flux at short wavelengths,
resulting in, e.g., $M_{F336W}$ magnitudes that are too bright by $\gta 0.12$
mag and UV-optical colours that are too blue by similar amounts.

In this investigation, isochrones from Paper I are applied to the {\it HST} UV
Legacy observations of six GCs that span the range in [Fe/H] from $\sim -0.7$ to
$\sim -2.3$.  The two main goals of this project are (i) an evaluation
of the quality of our BCs and stellar models in both an absolute and a
systematic sense, and (ii) an improved understanding of the chemical properties
of the selected clusters.  As we consider metal abundance mixtures that span
close to the maximum range in CN abundances, at a given value of [CNO/Fe], as 
well as very wide variations in the abundances of C, N, and O, our focus is on
comparisons of isochrones with the overall spreads in colours along the MS and
lower RGB in observed CMDs.  We consider this work to be a necessary first step
towards the greater goal of evaluating the absolute light element abundances of
the sub-populations of stars that have been revealed by chromosome maps.
Accordingly, no attempt is made in this study to analyze such maps.
(As an understanding of our results relies quite heavily on the material
presented in Paper I, it is
important that Paper I be read prior to this study.)

\section{Metal Abundance Mixtures and Bolometric Corrections}
\label{sec:bc}

This investigation is concerned with just a subset of the cases considered in
Paper I; specifically, those listed in Table~\ref{tab:t1}.  The names that have
been given to the different mixtures, which are listed in the first column,
describe the abundance variations and BCs that they represent.  All of them
begin with ``{\tt a4}" to indicate that they assume [$\alpha$/Fe] $= 0.4$ for
the $\alpha$ elements, while the subsequent letters identify the the main
abundance difference that distinguishes each mixture. Thus,``{\tt CN}" indicates
reduced C and increased N abundances consistent with CN-cycling, while ``{\tt
CNN}"  is similar to ``{\tt CN}" except that even higher N is assumed, implying
a larger value of C$+$N$+$O.  Likewise, ``{\tt a4ON}" and ``{\tt ONN}" are
indicative of mixtures that would be produced by efficient ON-cycling; the
latter differs from the former only in assuming a higher abundance of nitrogen.
The others allow for enhanced C by 0.4 dex ({\tt xC\_p4}), enhancements in both
C and O ({\tt xCO}), or increased O abundances by 0.2 or 0.4 dex ({\tt xO\_p2}
or {\tt xO\_p4}, respectively).
  
Whereas the previous generation of MARCS models (\citealt{gee08}) assumed the
solar abundances given by \citet{gas07}, the model atmospheres and improved
synthetic spectra that were computed for Paper I adopted the solar mixture of
the metals reported by \citet{ags09}.  When the abundances of the $\alpha$
elements are increased by 0.4 dex, we obtain the {\tt a4s08} and {\tt a4s21}
mixtures, which have ``{\tt 08}" or ``{\tt 21}" in their names to indicate, in
turn, the 2008 or 2021 MARCS models.  For both of these cases, and all others
that are listed in Table~\ref{tab:t1}, numerical values of the abundances of
several elements are provided on the usual scale in which $\log\,N$(H) $= 12.0$.
(The abundances are given explicitly only for those elements for which the
effects of altered abundances on BCs have been investigated.  Although there
are differences in the assumed He abundances between the {\tt a4s08} and {\tt
a4s21} mixtures, as indicated in the table, Paper I has demonstrated that they
affect the BCs at the level of only a few thousandths of a magnitude.)  For
the other cases (i.e., {\tt a4CN} $\ldots$ {\tt a4xO\_p4}), the changes in the
abundances relative to the standard [$\alpha$/Fe] $= 0.4$ mixture are specified
in dex along with upward- or downward-pointing arrows to indicate, in turn, 
enhanced or reduced abundances.  (The effects on BCs of differences in the
assumed microturbulent velocity, $v_T$, as noted in Table~\ref{tab:t1}, are
discussed in Paper I.)

As reported by \citet{pcs09} and \citet{cmp13}, isochrones on the theoretical 
plane are not affected by variations in C:N at constant C$+$N$+$O;
consequently, one can simply apply the BCs that are calculated for mixtures
with different C:N ratios to the isochrones for the {\tt a4s21} abundances of
C, N, and O (scaled to the [Fe/H] values of interest).  Paper I confirmed and
extended this result to stellar models of very low mass, though it also showed
that the $\teff$s of lower main-sequence (LMS) stars depend on the abundance 
of oxygen (and carbon, but to a lesser extent).  It is therefore necessary, if
one is interested in LMS stars, to take into account the effects of assumed C
and O abundance variations on both the model temperatures and the BCs that are
used to predict the magnitudes and colours of stars.  In this investigation,
Victoria-Regina (V-R) isochrones for [$\alpha$/Fe] $= 0.4$ have been transformed
to observed CMDs using the {\tt a4s21} BCs to represent CN-weak stars and the
{\tt a4CN} BCs to represent CN-strong stars.  Similarly, we have applied the
{\tt a4xO\_p2} or {\tt a4CNN} BCs to stellar models for [O/Fe] $= 0.6$, with
[$m$/Fe] $= 0.4$ for the other $\alpha$ elements, to represent CN-weak and
CN-strong stars that have higher N (and hence higher C$+$N$+$O) abundances.
Even though it was not necessary to compute grids of stellar models for the
{\tt a4CN} and {\tt a4CNN} mixtures, they were generated for these and all of
the other ``{\tt a4}" mixtures that are listed in Table~\ref{tab:t1} using the
same code that is described by \citet{vbd12}.

Paper I has already provided quite a detailed description of the procedure that
is used used to transpose isochrones from the theoretical plane to the various
CMDs, but it is worthwhile to include a brief summary of what was done here.  To
keep the total computational effort at a manageable level, model atmospheres,
synthetic spectra, and BCs were generated only for quite a sparse grid of
parameter values; specifically, $\log\,g = 1.0, 2.0, 3.0, 4.0, 4.5, 4.7, 5.0$, and 
5.3, with 4--6 $\teff$ values at each gravity, for each of three metallicities
([Fe/H] $= -2.5, -1.5$, and $-0.5$).  The grid values of $\teff$ were determined
from a consideration of isochrones for the selected [Fe/H] values.  Tables were
produced that contain not only these BCs (for each of the metal abundance
mixtures), but also those based on 2008 MARCS spectra at the same parameter
values; the latter were given the name {\tt a4s08} and derived from the
transformation tables provided by \citet[hereafter CV14]{cv14}.  These tables
make it possible to evaluate the differences in the BCs between any of the
``{\tt a4}" tables, in turn, and the reference {\tt a4s08} dataset at common
values of $\teff$, $\log\,g$, and [Fe/H].  Indeed, for a given isochrone, the
BCs for the predicted temperatures at each of the grid values of $\log\,g$ can
be evaluated, and the resultant $\delta$(BC) values as a function of $\log\,g$
fitted by cubic splines for subsequent interpolations.

The transformation of isochrone luminosities and temperatures to magnitudes and
colours involves the following steps.  First, the BCs for the selected filters
are derived from the CV14 tables, as they are provided for much finer spacings
of $\teff$, $\log\,g$, and [Fe/H] than the BCs that have been computed for this
project.  Second, splines are fitted to the differences in the BCs for the
selected mixture of the metals and the subset of the CV14 results that are
contained in the {\tt a4s08} tables.  Third, the splines are interpolated to
yield the $\delta$(BC) values at all intermediate values of $\teff$ and
$\log\,g$ along the isochrone (or extrapolated if $\log\,g < 1.0$ or $> 5.0$).
Finally, the resultant $\delta$(BC) values are applied to the
isochrone that was generated in the first step of this process.  Three-point
(quadratic) interpolation is used to derive the transformations for any
metallicity within the range $-2.5 \le$ [Fe/H] $\le -0.5$.  (Plots that
illustrate the spline fits to the $\delta$(BC) values and comparisons of
isochrones that are generated in the first and fourth steps of this procedure
are provided in Paper I.)

In what follows, the names listed in the first column of Table~\ref{tab:t1} are
used to refer to the metal abundance mixtures, to the BCs, or to the isochrones
for those mixtures.  

\section{Application of Stellar Models to Globular Cluster CMDs}
\label{sec:gc}

\begin{figure*}
\includegraphics[width=\textwidth]{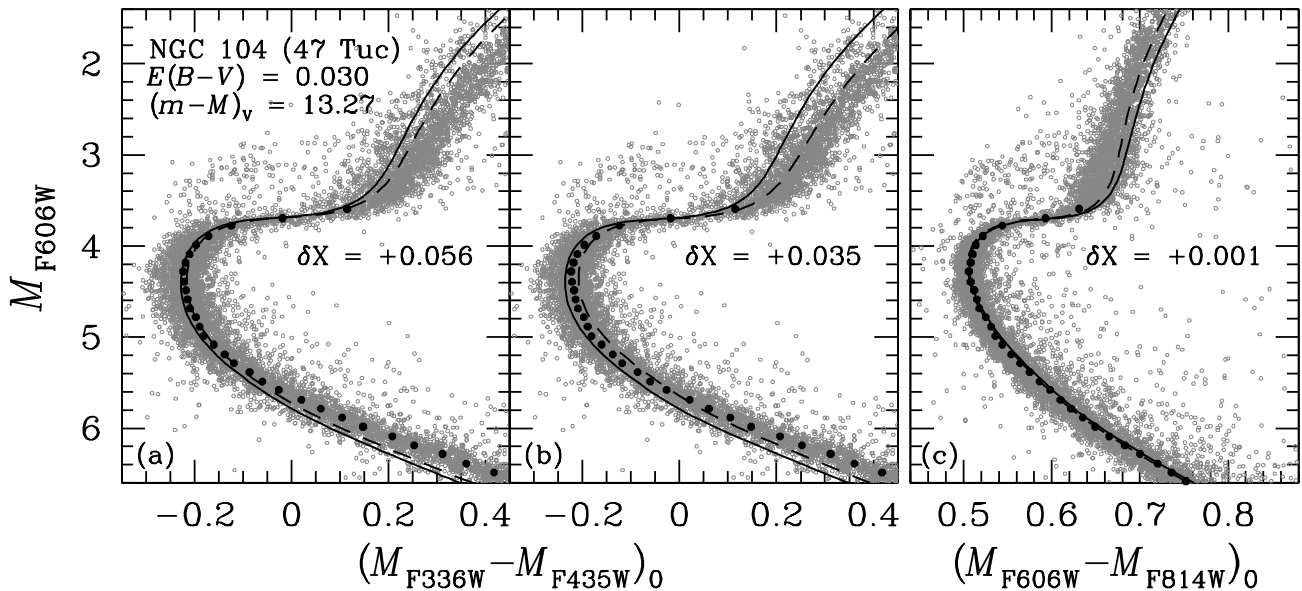}
\caption{{\it Panel (a):} Fit of 12.5 Gyr isochrones for [Fe/H] $= -0.70$,
$Y= 0.27$, and [$\alpha$/Fe] $= 0.4$ to observations of 47 Tuc (from NLP18),
on the assumption of the indicated reddening and apparent distance modulus.
(47 Tuc is the only GC considered in this study for which NLP18 provide ACS
$F435W$, instead of WFC3 $F438W$, magnitudes.)  The isochrones
have been transformed to the observed plane using the {\tt a4s21}
and {\tt a4CN} BCs, resulting in the solid and dashed curves, respectively.  In
order to fit the median fiducial sequence in the vicinity of the TO, the
specified value of $\delta\,X$ was applied to the isochrone colours (see the
text).  {\it Panel (b):} As in panel (a) except that the isochrones assume an
age of 12.0 Gyr and a higher oxygen abundance by 0.2 dex; consequently, the
solid and dashed curves employed the BCs for, in turn, the {\tt a4xO\_p2} and
{\tt a4CNN} mixtures (see Table~\ref{tab:t1}).  {\it Panel (c):} As in panel (b)
except that the isochrones have been fitted to $F606W, F814W$ photometry.}
\label{fig:f1}
\end{figure*}

The photometric data for all of the GCs considered in this paper were released
into the astronomical community by NLP18 via the website that they
have provided.  We have opted to use
their ``Method 1" photometry, which is preferred for the upper MS and more
evolved stellar populations.  Moreover, the CMDs were limited to stars with
membership probabilities $\gta 98$\%, photometric errors $< 0.02$ mag, and
quality of fit ({\tt QFIT}) parameters $\gta 0.99$, though these criteria were
relaxed or tightened somewhat in order that the number of selected stars was
sufficient to produce well-defined sequences from the upper MS to the lower
red-giant branch (RGB).  The MS and TO observations were sorted into 0.1 mag
bins in $m_{F606W}$, and median fiducial points were determined for each bin.
By fitting isochrones to the median fiducial sequence instead of the entire
distribution of the individual stars in the vicinity of the TO, subjective
errors associated with the determination of the best estimate of the TO age that
corresponds to an adopted distance modulus are essentially eliminated.

We have deliberately selected several GCs for this study that have been the
subject of recent investigations in which the cluster distance moduli were
obtained either from fits of zero-age horizontal branch (ZAHB) loci to the
observed HB populations (e.g., \citealt[hereafter VBLC13]{vbl13}, 
\citealt{vdc16}) or from full simulations of the latter (e.g., \citealt{dvk17}).
In all such studies, the HB models provide fully consistent extensions of the 
Victoria-Regina isochrones (\citealt{vbf14}) to the core He-burning phase.  To
ensure that the results obtained in this study are compatible with those of
previous investigations, we have adopted the same, or very similar, values of
$(m-M)_V$, $E(B-V)$, and [Fe/H].  These choices can be justified by the fact
that the Victoria-Regina isochrones satisfy the constraints provided by, e.g.,
solar neighborhood Population II stars (\citealt{vcs10}), the morphologies of
GC CMDs (see the aforementioned papers as well as, e.g., \citealt{cgk16},
\citealt{sdf18}), and the properties of cluster binaries (e.g., \citealt{bvb17},
\citealt{vd18}).  Furthermore, the distance moduli that have been derived from
the HB models agree rather well with those inferred from the RR Lyrae standard
candle (e.g., VBLC13, \citealt{dvk17}).\footnote{As a further check of the 
ZAHB-based distance scale, we compared the true distance moduli that were
derived by VBLC13 from fits of ZAHB models to the HB populations of 43 GCs
with the results reported by \citet{bv21}, which are based on Gaia EDR3
parallaxes and their compilation of literature values.  For 27 (63\%) of the
clusters, the differences in $(m-M)_0$ are $\le 0.05$ mag; only for 6 of the
GCs are the differences $> 0.10$ mag.}   

Our focus is on several of the CMDs that can be constructed from $F336W$,
$F438W$, $F606W$, and $F814W$ photometry.  Because the MARCS spectra do not
extend sufficiently far into the UV, it was not possible to calculate BCs for
the $F275W$ filter; consquently, we are unable to fit isochrones to the
available $F275W$ observations.  It should noted as well that, although the CMD
plots specify the values of $E(B-V)$ and $(m-M)_V$ that have been adopted for
each GC, the excess in a given $\zeta - \eta$ colour or the extinction $A_\zeta$
has been calculated from the {\it nominal} $E(B-V)$ value using the values of
$R_\zeta = A_\zeta/E(B-V)$ given by CV14 (their Table A1).  Thus, for instance,
the difference between the apparent and the absolute $F606W$ magnitudes, which
are plotted along the $y$-axis, can be easily calculated from $(m-M)_{F606W} =
(m-M)_V + (R_{F606W} - R_V)E(B-V)$.

\subsection{NGC\,104 (47 Tuc)}
\label{subsec:tuc}

\begin{figure*}
\includegraphics[width=\textwidth]{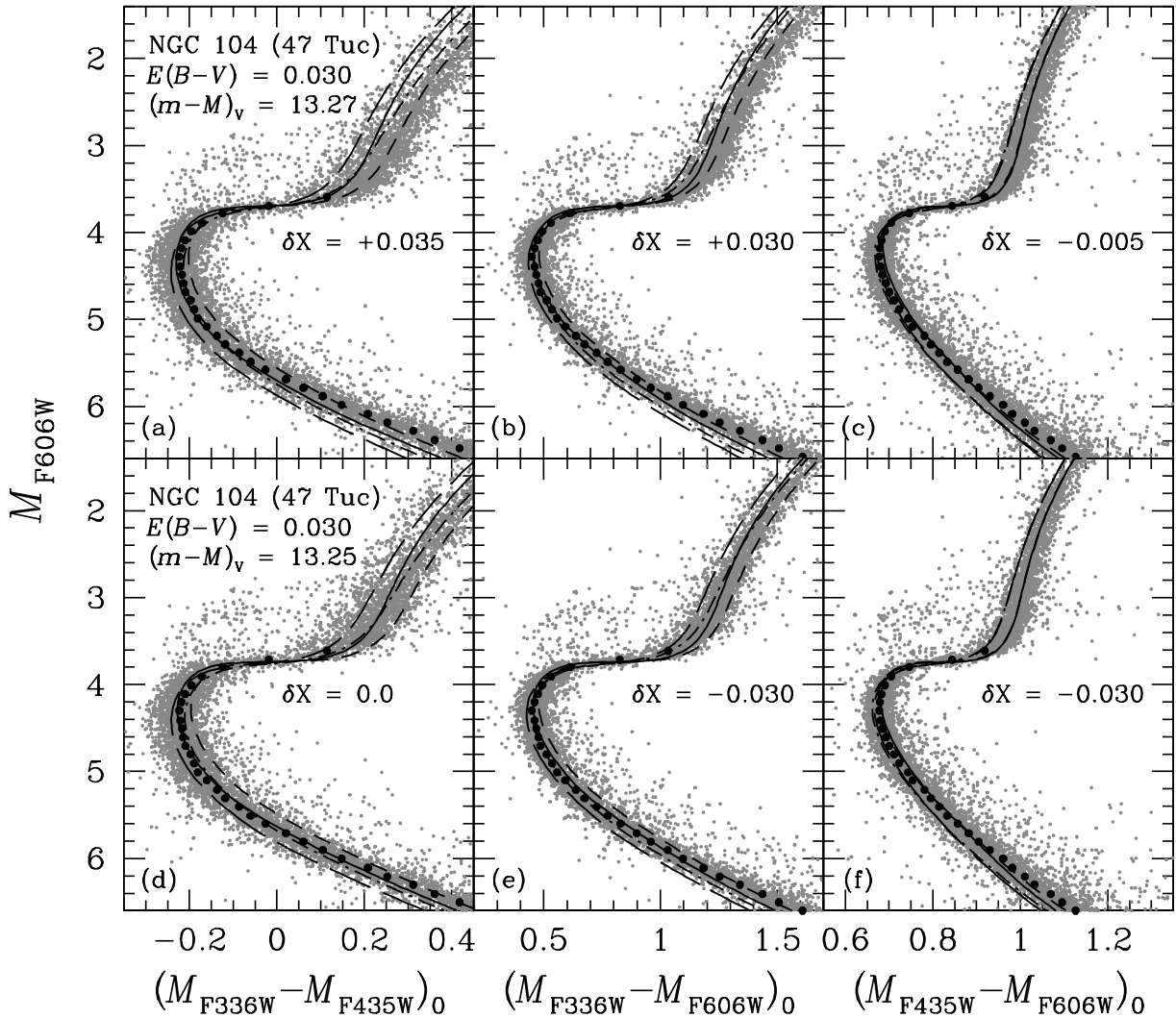}
\caption{{\it Panels (a)--(c):} As in the middle and right-hand panels of the
previous figure, except that the isochrones represented the solid and dashed
curves assume $Y = 0.25$, while those plotted as the long-dashed and dot-dashed
loci assume $Y = 0.30$. {\it Panels (d)--(f):} As in panels (a)--(c), except
that the isochrones assume [Fe/H] $= -0.60$ and an age of 11.9 Gyr.}
\label{fig:f2}
\end{figure*}

With an absolute integrated visual magnitude of $-9.42$ according to the 2010
edition of the \citet{har96} catalogue, 47 Tuc is one of the most massive GCs
in the Milky Way.  Its basic parameters appear to be quite well established.
\citet{bvb17} concluded from an examination of the available evidence that the
best estimate of the foreground reddening is $E(B-V) = 0.030 \pm 0.01$, and
their analysis of an eclipsing binary member known as V69 suggested a
preference for [Fe/H] $ = -0.70$, [$\alpha$/Fe] $= 0.4$, [O/Fe] $\approx 0.6$
and $Y \approx 0.25$.  (By comparison, the latest spectroscopic survey gives
[Fe/H] $= -0.76$; see \citealt[hereafter CBG09]{cbg09a}.)  V69 could well be a
member of a helium-poor population given that simulations of the cluster HB
stars by \citet{dvk17} have indicated that the star-to-star variation of the
initial He abundance in 47 Tuc is $\delta\,Y_0 \approx 0.03$, with a mean value
close to 0.27.  The same simulations yield $(m-M)_V = 13.27$ if the metallicity
is taken to be [Fe/H] $= -0.70$ and the faintest HB stars have $Y_0 = 0.257$.
This is in very good agreement with the determination of $(m-M)_V = 13.30$ by
Brogaard et al.~from the binary V69, and with the true modulus derived by
\citet{crc18}, $(m-M)_0 = 13.24 \pm 0.06$, if $A_V \lta 0.09$\ mag.  (As the
distance reported by Chen et al.~was obtained from Gaia DR2 parallaxes with a
correction to the zero-point of the parallax scale based on observations of
quasars and background stars in the Small Magellanic Cloud, it is independent
of stellar models.)

In contrast with all other GCs considered in this study, NLP18 provide {\it HST}
Advanced Camera for Surveys (ACS) $F435W$ photometry for 47 Tuc instead of WFC3
$F438W$ data.  However, this is of little consequence, as our isochrones indicate
that predicted colours involving these magnitudes (e.g., $M_{F336W}-M_{F435W}$
versus $M_{F336W}-M_{F438W}$) are nearly identical once a small zero-point
difference ($\approx 0.01$ mag) between the $F438W$ and $F435W$ BCs is taken
into account.  Of the various CMDs considered here, one can
anticipate that the $(M_{F336W}-M_{F435W})_0,\,M_{F606W}$ diagram will
provide the most challenging test of stellar models because the selected colour
involves the UV filter, $F336W$.  According to Figure~\ref{fig:f1}a, isochrones
for [$\alpha$/Fe] $= 0.4$ that are relevant for CN-weak and CN-strong stars (the
solid and dashed curves, respectively), have to be adjusted by quite a large
amount in the horizontal direction (0.056 mag) in order to match the observed
TO colour.\footnote{NLP18 tabulate the uncertainties in the photometric
zero-points that are used to calibrate instrumental magnitudes in the VegaMag
system.  For most clusters, they are at the level of 0.01--0.02 mag for $F275W$
and the other WFC3 filters, implying possible colour errors of
$\sim 0.014$--0.028 mag. This could well be responsible for part of the
$\delta$\,X offsets that are needed to reconcile predicted and observed TO
colours, but they may also be the result of errors in the model $\teff$ scale
and in the BC--$\teff$--[Fe/H] relations, both of which are hard to quantify,
as well as to errors in the assumed cluster properties (i.e., metallicities,
distances, and foreground reddenings).}  Furthermore,
even when this offset is applied to the models, the
isochrones fail to reproduce the CMD locations of the lower RGB stars.  (Note
that the adopted correction to the colours was chosen so that the isochrones
would straddle the median fiducial sequence in the vicinity of the TO; this was
done on purpose to reflect the fact that the distribution of CN strengths is
known to be strongly bimodal in 47 Tuc; see, e.g., \citealt[their
Fig.~3]{ccb98}.)

However, as shown in Fig.~\ref{fig:f1}b, isochrones for a higher abundance of
oxygen by 0.2 dex, as transformed to the observed plane using the {\tt
a4xO\_p2} BCs to represent CN-weak stars (solid curve) and the {\tt a4CNN} BCs
to represent CN-strong stars (dashed curve) provide significantly improved fits
to the observed CMD.  The colour offset, $\delta$\,X, has been reduced to 0.035
mag and the overlay of the model loci onto the cluster giants has improved,
though the stellar models are still too blue.  (Presumably both of these
discrepancies between theory and observations would be even smaller if
isochrones for [O/Fe] $> 0.6$ were fitted to the observations, but there is no
(other) evidence for such high oxygen abundances at [Fe/H] $\gta -0.7$.)  One of
the consequences of a 0.2 dex increase in the O abundance is a modest reduction
in the TO age to $\approx 12.0$ Gyr from $\approx 12.5$ Gyr (in panel a), if the
same apparent distance modulus is adopted.  

Fig.~\ref{fig:f1}c has been included here to show that the same isochrone loci
that appear in panel (b) provide superb fits to the $(M_{F606W}-M_{F814W})_0$
colours of the upper MS and TO stars in 47 Tuc without requiring a significant
adjustment to the predicted colours.  This
argues against the possibility that the differences between the predicted and
observed colours in Fig.~\ref{fig:f1}b are mainly due to problems with the model
$\teff$ scale.  As regards the discrepancies along the lower RGB: the relative
locations of the solid and dashed curves indicate that $F606W,\,F814W$
photometry is sensitive to the abundance of nitrogen.  Since our models for the
{\tt a4CNN} mixture assume [N/Fe] $= 1.13$, as compared with measured abundances
that range up to [N/Fe] $\sim +1.6$ (see \citealt[their Fig.~6]{bcs04};
\citealt[their Fig.~11]{cbs05}), computations for higher N abundances may do
a better job of reproducing the colours of the bluest giants.  Further work is
needed to investigate this possibility.

The difficulties that are apparent in Fig.~\ref{fig:f1} can be alleviated to a
considerable extent by allowing for He abundance variations.  Helium mainly
affects the predicted gravities and temperatures of stars.  It has almost no
impact on the BCs at fixed values of [Fe/H], $\teff$, and $\log\,g$ ---
something that was verified in Paper I by comparing BCs for the same mixtures
of the metals, but different $Y$ (also see \citealt{gcb07}).  In fact, a 
star-to-star He abundance variation corresponding to $\delta\,Y \sim 0.05$,
which is on the high side for most GCs, is predicted to have almost no effect
on the $M_{F336W}-M_{F435W}$ colours of TO stars, while causing a spread of the
same colour along the MS and lower RGB by $\sim 0.04$--0.06 mag.  This is shown
in Figure~\ref{fig:f2}a, which plots 12.0 Gyr isochrones for the same
oxygen-enhanced abundances as in Fig.~\ref{fig:f1}b, but for $Y = 0.25$ and
0.30.  Along the giant branch, the horizontal separation between the solid and
long-dashed isochrones, or between the dashed and dot-dashed loci, is about
half of that predicted by the models that represent CN-weak and CN-strong stars
(e.g., the solid and dashed isochrones) at constant $Y$.  Figs.~\ref{fig:f2}b
and~\ref{fig:f2}c show that the same isochrones provide comparable fits to
other CMDs that can be constructed from $F336W$, $F435W$, and $F606W$ photometry.

Clearly, variations in both $Y$ and CN should be taken into account when 
attempting to explain the observed widths, at a given magnitude, of both
the MS and the lower RGB of 47 Tuc.  In fact, our models are able to explain
the {\it thicknesses} of the principal photometric sequences quite well if
$\delta\,Y \sim 0.05$.  The main difficulty with the stellar models is that they
are generally too blue along the lower RGB (in all three of the CMDs that have
been plotted).  One possible way of improving the fits to the observations is to
adopt a higher metallicity (i.e., [Fe/H] $> -0.70$), which has been found in
some spectroscopic studies (see, e.g., \citealt{cgb04}, \citealt{wcs06},
\citealt{jmp15}).  As shown in Figs.~\ref{fig:f2}d--f, isochrones
for a higher [Fe/H] value by 0.1 dex provide better fits to the cluster giants
by virtue of being cooler and therefore redder; indeed, the region enclosed by
the bluest and reddest isochrones contains most of the MS, TO, and giant-branch
stars.  These models also require less of a redward shift of their 
$M_{F336W}-M_{F435W}$ colours to match the observed TO, though similar or larger
colour adjustments are needed in the case of the other colours.  It would 
therefore appear that isochrones for an intermediate metallicity (say, [Fe/H]
$\approx -0.65$), together with a higher nitrogen abundance than in our current
models, would provide the best overall fits to the various CMDs.  (Note that
ZAHB models for [Fe/H] $= -0.60$ yield a smaller value of $(m-M)_V$ by 0.02 mag
and that the best estimate of the corresponding TO age is just slightly reduced
to 11.9 Gyr.)

\begin{figure}
\includegraphics[width=\columnwidth]{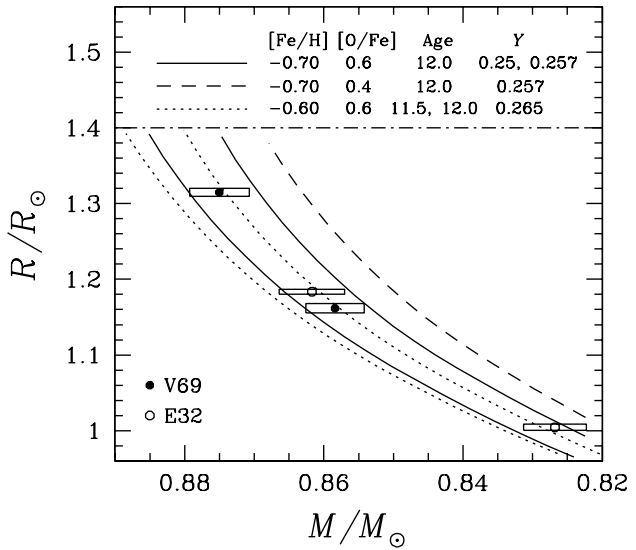}
\caption{Comparison of the predicted relations between mass and radius from
several isochrones that have the indicated ages and chemical abundances with
the properties of the eclipsing binaries V69 and E32 in 47 Tuc (filled and open
circles, respectively, and the associated error boxes).  Their masses and radii
were taken from the study by \citet{tud20}.  The adopted value of [O/Fe] is
given explicitly; all other $\alpha$-elements are assumed to have [$m$/Fe]
$= 0.4$.}
\label{fig:f3}
\end{figure}

Fortunately, member eclipsing binary stars provide important constraints on 
the chemical composition of 47 Tuc.  \citet{tud20} have recently reduced (albeit
only slightly) the uncertainties associated with the physical properties of V69,\
as compared with the findings of \citet{bvb17}, and they have determined precise
masses and radii for a second binary, E32, to within 0.55\% and 0.40\%,
respectively.  Their results for the two binaries are compared with the $M$--$R$
relations from several isochrones in Figure~\ref{fig:f3}.  The solid curve
located to the left of the binary components, which predicts higher masses at a
given radius, represents the same isochrone (for $Y = 0.25$) that was fitted to
the 47 Tuc photometry in Figs.~\ref{fig:f2}a--c, while the other one assumes
$Y = 0.257$.  These results show that a relatively low He abundance ($0.250
\lta Y \lta 0.257$) is required by the binary constraint if 47 Tuc has [Fe/H]
$= -0.70$, [O/Fe] $= 0.6$ (with [$m$/Fe] $= 0.4$ for the other $\alpha$
elements), and an age near 12.0 Gyr.  The assumption of a higher O abundance
(and hence higher [CNO/Fe]) would shift the predicted $M$--$R$ relations further
to the left, in which case, the binaries would favor a somewhat higher $Y$.
(This would actually be more consistent with expectations for a relatively
metal-rich GC if $\delta\,Y/\delta\,Z \gta 1$ as the result of chemical
evolution since the Big Bang.)

By comparison, the dotted curves represent isochrones for the same [Fe/H] 
value ($-0.60$) and the oxygen-enhanced mixture of the metals that were plotted
in Figs.~\ref{fig:f2}d--f, though we have assumed $Y = 0.265$ for
the He abundance so that a 12.0 Gyr isochrone provides a good fit to the
properties of the binary components.  The dotted curve that lies furthest to 
the left in this plot shows the effect on the $M$--$R$ relation of a reduction
in the age by 0.5 Gyr.  The horizontal shift between the two dotted loci is
apparently nearly the same as that caused by $\delta\,Y \approx 0.007$, which
is responsible for the separation at constant radius between the two solid
curves.

The dashed curve in Fig.~\ref{fig:f3} is of particular importance because it
has been derived from a 12.0 Gyr isochrone for [Fe/H] $=-0.70$, $Y = 0.257$,
and [O/Fe] $=$ [$\alpha$/Fe] $= 0.4$.  It is clearly problematic insofar as
it predicts masses for the binary at the observed radii that are too low.  (In
fact, this isochrone would match the observed TO luminosity only if a larger
distance modulus by $\sim 0.07$ mag were adopted.  Smaller values of $(m-M)_V$
would imply higher ages and reduced masses.)  This is the reason why
\citet{bvb17} favoured a higher O abundance if the metallicity of 47 Tuc is
[Fe/H] $= -0.70$.  Although the distance modulus uncertainty permits some
flexibility in the fits to the mass-radius diagram, we have been able to obtain
a consistent interpretation of both the WFC3 CMDs of 47 Tuc and its eclipsing
binaries on the assumption of [Fe/H] $\gta -0.70$ and [O/Fe] $\sim 0.6$.  
The small discrepancies that remain between predicted and observed colours, in
particular, may well be reduced if we were to adopt alternative choices for
the abundances of the CNO elements and/or though further improvements to
the computation of synthetic spectra and BCs --- but this must be left for
future work to determine.  Perhaps the main point of our analysis is that
star-to-star variations in the abundances of just the three elements, He, C, 
and N appear to be able to account for most of the observed colour spreads 
along the MS and RGB of 47 Tuc.

In concluding this section, some discussion is warranted concerning the fact
that the isochrones which were fitted to the CMDs of NGC$\,$6496 in Paper I
required much smaller colour offsets ($\lta 0.01$ mag) than those obtained for
47 Tuc, despite having a similar metallicity to within $\sim 0.2$ dex.  We
suspect that the most likely explanation of this difference is that the
photometry of NGC$\,$6496 was fitted by models that assumed [$\alpha$/Fe] $= 0.4$,
which is probably too high.  As this cluster has [Fe/H] $\gta -0.5$ (CBG09), it
would be expected to have [$\alpha$/Fe] $\sim 0.25$ if it lies close to the
standard relation between [$\alpha$/Fe] and [Fe/H] that has been derived for
field Population II stars (e.g., \citealt{eag93}, \citealt{fuh08}).  Indeed, we
have verified, using V-R isochrones and the BCs provided by CV14, that the models
for lower abundances of the $\alpha$ elements by $\sim 0.15$ dex would require
redward colour shifts that are similar to those found for 47 Tuc.  However,
differences in the C$+$N$+$O abundance between NGC$\,$6496 and 47 Tuc could also
affect how well stellar models are able to reproduce their respective TO colours.

\subsection{NGC\,6362}
\label{subsec:6362}
  
\begin{figure}
\includegraphics[width=\columnwidth]{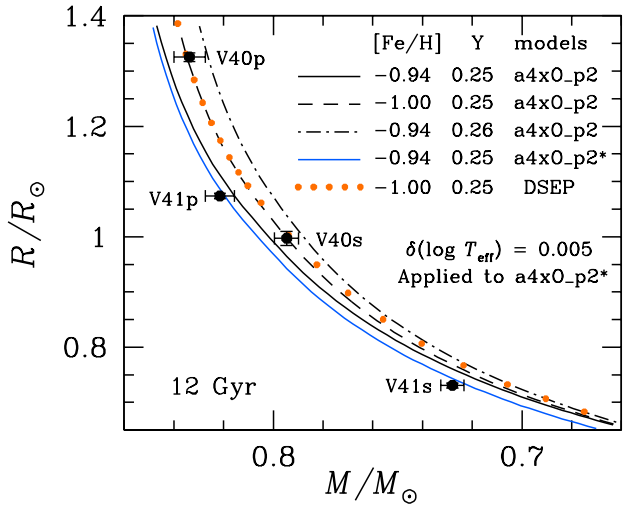}
\caption{Comparison of the $M$--$R$ relations predicted by 12 Gyr isochrones that
have the indicated chemical abundances with the properties of the eclipsing
binaries (V40 and V41) in NGC\,6362 (filled circles and error bars) as derived
by \citet{ktd15}.  The only difference between the solid curves in blue and
black is that the predicted temperatures along the former were arbitrarily
increased by $\delta\log\teff = 0.005$, resulting in smaller radius at fixed
uminosities.  An isochrone from the Dartmouth (DSEP) database (\citealt{dcj08})
for very close to the same metal abundances as the {\tt a4xO\_p2} models (see
the text) has also been plotted (small filled circles in orange).}
\label{fig:f4}
\end{figure}

\begin{figure*}
\includegraphics[width=\textwidth]{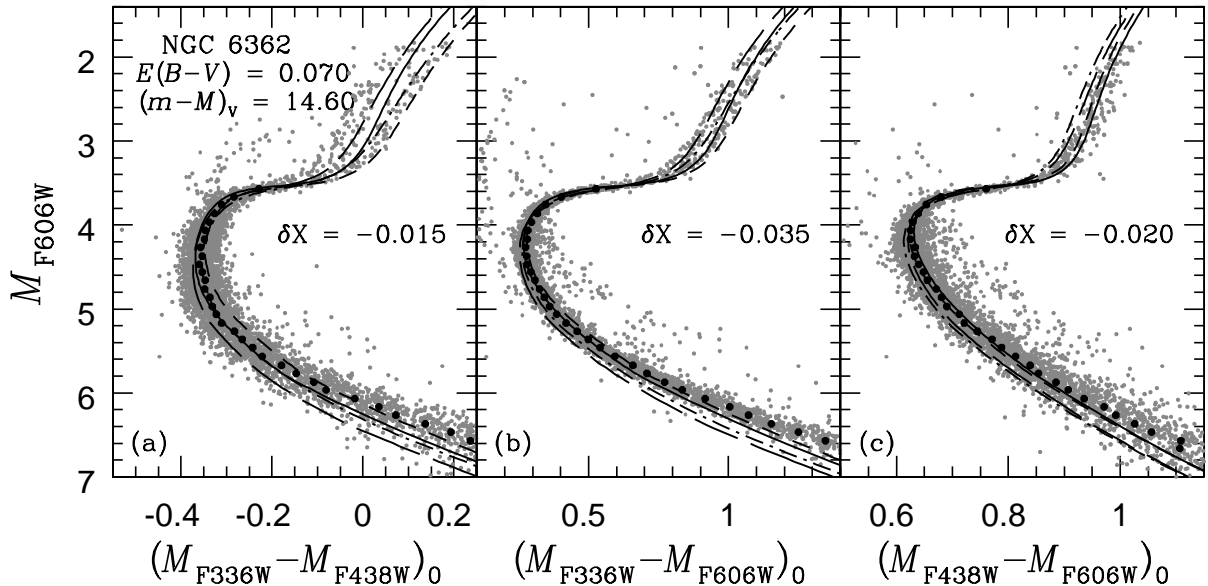}
\caption{Similar to the top (or bottom) row of panels in Fig.\ref{fig:f2}; in
this case, WFC3 observations of NGC\,6362 (from NLP18) have been fitted by 11.8
Gyr isochrones for [Fe/H] $= -0.94$, $Y= 0.25$, [$\alpha$/Fe] $= 0.4$, and
[O/Fe] $= 0.6$ employing either the {\tt a4xO\_p2} or {\tt a4CNN} BCs (solid
and dashed curves, respectively).  Otherwise identical isochrones, but for $Y =
0.30$ are represented, in turn, by the long-dashed and dot-dashed loci.}
\label{fig:f5}
\end{figure*}
 
NGC\,6362 should be a particularly good GC for the testing and calibration of
stellar models for [Fe/H] $\sim -1.0$ because it contains a large number of
RR Lyrae variables, as well as eclipsing binaries with well determined
properties.  In their recent study of this system, \citet{vd18} found that these
constraints could be satisfied quite well if NGC\,6362 has $E(B-V) = 
0.07$--0.08, which is consistent with the reddenings derived from dust maps
(\citealt{sfd98}, \citealt{sf11}), and an apparent distance modulus in the range
of $(m-M)_V = 14.56$--14.60, if the cluster has $-1.0 \lta$ [Fe/H] $\lta -0.85$.
Although many spectroscopic studies have found metallicities somewhat below
$-1.0$, including the surveys by \citet{ki03} and CBG09, the binary stars appear
to preclude such low values if they have $Y_0 \gta 0.25$ and [O/Fe] $\lta 0.6$.
Simulations of the cluster HB that were presented in the same paper by 
VandenBerg and Denissenkov indicated that the initial He abundance varies by
$\delta\,Y_0 \approx 0.03$, with $\langle Y_0\rangle \approx 0.26$.  

To illustrate the difficulties presented by the eclipsing binaries for
[Fe/H] $\lta -1.0$, we have plotted in Figure~\ref{fig:f4} the masses and radii
of the binaries V40 and V41, together with the uncertainties in these
properties, from the study by \citet{ktd15}.  Superimposed on the observations
are the predicted mass-radius relations from 12 Gyr isochrones for the
{\tt a4xO\_p2} mixture, assuming the indicated values of $Y$ and [Fe/H].  (An
age close to 12 Gyr is expected if NGC\,6362 has an apparent distance modulus
close to $(m-M)_V = 13.60$; see below.)  The solid curve in black shows that the
properties of the primary of V41 can be matched by isochrones for $Y = 0.25$,
which should be very close to the minimum possible value because it is
approximately the primordial He abundance (\citealt{cfo16}), only if [Fe/H]
$\gta -0.94$ is assumed.  (It is much easier to accommodate either a somewhat
lower metallicity or a higher He abundance in the case of V40; note its location
relative to the dashed and dot-dashed curves.)  If the adopted metallicity is
decreased by as little as 0.06 dex, the corresponding $M$--$R$ relation (the
dashed curve) is well outside the $1\,\sigma$ error box of V41p.  Although not
shown, nearly the same relation as the dashed curve is obtained if the O
abundance is decreased by 0.2 dex (to [O/Fe] $= 0.4$).  These results provide
ample justification for adopting [O/Fe] $= 0.6$ in the fits of isochrones to
the cluster CMDs to be presented shortly. 

\begin{figure}
\includegraphics[width=\columnwidth]{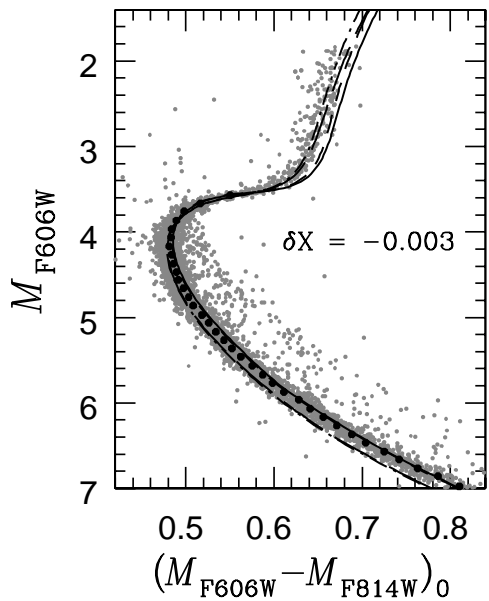}
\caption{As in the previous figure, except that the isochrones have been
fitted to WFC3 $F606W,\,F814$ observations of NGC\,6362.}
\label{fig:f6}
\end{figure}

The small filled circles in orange, which represent the $M$--$R$ relation
predicted by an isochrone from the Dartmouth database (\citealt{dcj08}) for the
same age, metallicity, and He abundance as the dashed curve, but for
[$\alpha$/Fe] $= 0.4$, demonstrate that our models agree quite well with the
results of a completely independent stellar evolution code.  Although the
Dartmouth isochrones adopted the solar abundances reported by \citet{gs98}
as the reference mixture of the metals, it turns out that their models for
[$\alpha$/Fe] $= 0.4$ assume nearly the same C$+$N$+$O abundance as our
isochrones for [O/Fe] $= 0.6$ and [$m$/Fe] $= 0.4$ for the other $\alpha$
elements; in fact, the two mixtures have the same value of $\log$(C$+$N$+$O) to
within 0.01 dex.   This explains why the dashed and orange loci are so similar.
Since both of these curves predict masses, at the derived radii of V41, that are
significantly lower than the observed masses of this binary, it would appear to
be quite a robust result that NGC$\,$6362 has [Fe/H] $\gta -1.0$ and a high
value of [CNO/Fe].

Interestingly, V41 appears to lie along a different mass-radius relation than
V40, implying that the helium abundances of the two binaries differ by
$\delta\,Y \sim 0.01$.  However, it is clearly much more difficult to obtain
a consistent interpretation of V41 than of V40, whose components can be fitted 
equally well by the same $M$--$R$ relation.  Our isochrones apparently
predict radii at the observed mass of V41s that are too large.  This
could be telling us that the temperatures of our stellar models are too high.
If the {\tt a4xO\_p2} isochrone for [Fe/H] $=-0.94$, $Y = 0.25$,
and [O/Fe] $= 0.6$ (the solid curve in Fig.~\ref{fig:f4}) is arbitrarily shifted
to higher $\teff$s by $\delta\log\teff = 0.005$ (approximately 70~K at $\teff =
6000$~K), it provides a satisfactory fit to the properties of both components of
V41.  This is illustrated by the solid curve in blue.  (A shift of the model 
loci in Fig.~\ref{fig:f4} to smaller radii would tend to increase the cluster He 
abundance as inferred from its binaries.)  The possibility of temperature scale
errors should be kept in mind as they are bound to be present at some level. 

%

Fits of 11.8 Gyr isochrones for [Fe/H] $= -0.94$, $Y = 0.25$, [$\alpha$/Fe] $=
0.4$, and [O/Fe] $= 0.6$ to the WFC3 CMDs of NGC\,6362 are shown in
Figures~\ref{fig:f5} and~\ref{fig:f6}; the {\tt a4xO\_p2} and {\tt a4CNN} BCs
were used to generate the solid and dashed loci, respectively.  The isochrone
represented by long dashes assumes $Y = 0.30$ but is otherwise identical to the
solid curve. The clear separation of the lower RGB stars into two sequences is
especially striking in the left-hand panel of Fig.~\ref{fig:f5}.  This could be
the manifestation of a strong bimodality in CN strengths, at least in part.
Indeed, this is suggested by the isochrones that have been superimposed on the
observations.  

However, although models that allow for variations in $Y$ and CN
strengths are able to match the full width of the RGB, at a given $F606W$
magnitude, in panel (b), they are less successful in this regard in the other
two panels.  Moreover, they fail to explain the bluest giants in panel (a), or
the reddest ones in panel (c), which cannot be attributed to errors in the model
$\teff$ scale or to the assumed He abundances because they cannot cause a
redward offset of the models in some CMDs and blueward shifts in others.  (We
defer our discussion of possible causes of this difficulty to our analyses of
the M\,5 and M\,3 CMDs where the same discrepancies between theory and
observations are even more pronounced.)  On the other hand, aside from the
apparent differences between the predicted and observed MS slopes, our
isochrones are able to reproduce the observed MS widths satisfactorily.  As in
the case of 47 Tuc, our isochrones provide good fits to the observed
$M_{F606W}-M_{F814W}$ colours of MS stars in NGC\,6362, though they are too
red along the giant branch (see Fig.~\ref{fig:f6}).\footnote{Because the fits of
isochrones to $F606W,\,F814W$ observations of all of the GCs considered in this
paper look so similar --- i.e., the isochrones reproduce the CMD locations of
MS and TO stars very well, with little or no offsets to the predicted colours,
and they are always too red along the giant branch by $\sim 0.03$--0.04 mag at
a given absolute magnitude --- we decided to include such plots in this paper
only for 47 Tuc and NGC\,6362.  Similar plots for the other clusters considered
in this investigation are qualitatively nearly identical and therefore do not 
add anything to our understanding.}  

Encouragingly, the $\delta\,X$ colour offsets that must be applied to the 
isochrones in order to match the observed TO colours are relatively small.  The 
fact that they are all negative (i.e., the isochrones must be adjusted to
bluer colours) does raise the concern that the adopted [Fe/H] value may be too
high, as the differences between the predicted and observed colours would be
less if lower metallicity, and therefore bluer, isochrones were fitted to the
observations.  Indeed, for most of the GCs considered in this study, the
predicted $M_{F336W}-M_{F438W}$ and $M_{F336W}-M_{F606W}$\ colours must be
{\it increased} by small amounts in order to reproduce the observed colours ---
though the isochrones that are fitted to their CMDs assume [Fe/H] values that
have stronger support from spectroscopy.  In fact, we would have found the same
thing for NGC\,6362 (i.e., small positive colour offsets) had we adopted, say,
[Fe/H] $= -1.07$ (CBG09).  However, if the cluster actually has a lower
metallicity by $\sim 0.1$ dex than we have assumed, consistent fits to both the
photometric data and the eclipsing binaries would presumably require a higher
C$+$N$+$O abundance than in our current stellar models (as already mentioned). 

\subsection{NGC\,5904 (M\,5)}
\label{subsec:m5}

\begin{figure*}
\includegraphics[width=\textwidth]{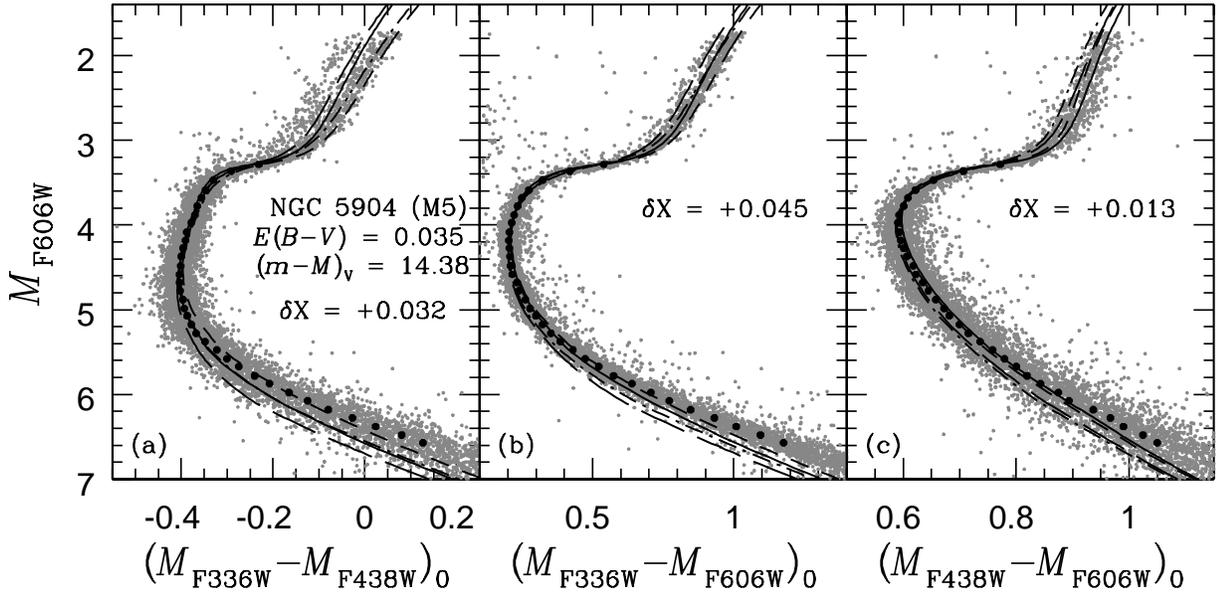}
\caption{Similar to the previous figure; in this case, 11.4 Gyr isochrones for
[Fe/H] $= -1.33$, $Y= 0.25$, [$\alpha$/Fe] $= 0.4$, and [O/Fe] $= 0.6$ have
been fitted to WFC3 observations of M\,5 (from NLP18) on the assumption
of the indicated reddening and apparent distance modulus.  The long-dashed and
dot-dashed loci represent isochrones for $Y = 0.30$ but are otherwise identical
to the solid and dashed curves, respectively.  Fits to $F606W,\,F814W$
photometry (not shown) look very similar to those given in Fig.~\ref{fig:f1}c;
in this case, the models match the MS and TO observations without requiring any
adjustments to the predicted colours.}
\label{fig:f7}
\end{figure*}

With about half of the metallicity of NGC\,6362, M\,5 is a suitable cluster
to consider as we extend our analyses to more metal-deficient systems in steps
of $\approx 0.3$ dex in [Fe/H].  Its basic parameters appear to involve rather
little controversy.  According to the spectroscopic survey by CBG09,
M\,5 has [Fe/H] $= -1.33$, and if ZAHB models for this metallicity and $Y =
0.25$ are fitted to the cluster HB, one obtains $(m-M)_V = 14.38$ (see, e.g., 
VBLC13).  The reddening appears to be close to $E(B-V) = 0.035$ insofar
as this estimate is within 0.003 mag of the foreground reddenings that are found
from the \citet{sfd98} and \citet{sf11} dust maps.  If 11.4 Gyr isochrones for
the same O- and $\alpha$-enhanced mixture of the metals that was adopted for
47 Tuc and NGC\,6362 are fitted to the WFC3 CMDs of M\,5, we obtain the
results shown in Figure~\ref{fig:f7}.  

\begin{figure*}
\includegraphics[width=\textwidth]{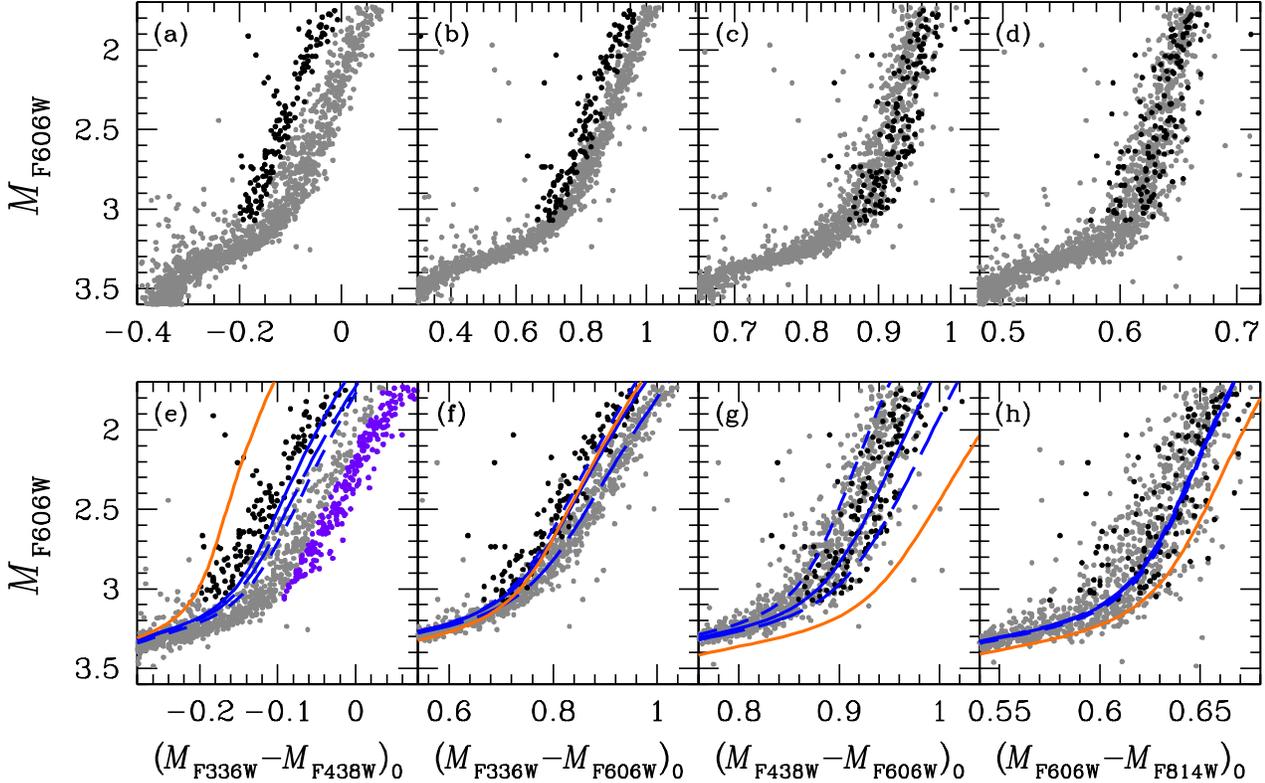}
\caption{{\it Panels (a)--(d)}: Magnified view of the observed subgiant and lower
RGB stars in M\,5 from the previous figure.  The stars represented by filled 
circles in black were selected on the basis of their $(M_{F336W}-M_{F438W})_0$
colours (left-hand panel), where they are clearly separated from the redder
giants.  The locations of the same stars in the other three CMDs are similarly
represented by black filled circles.  {\it Panels (e)--(h)}: the same CMDs from
the top row are superimposed by the lower RGB portions of isochrones for the
{\tt a4xO\_p2}, {\tt a4xC\_p4}, and {\tt a4xCO} mixtures (the dashed, solid,
and long-dashed loci in blue, respectively), along with an isochrone for the
{\tt a4s08C1} mix (see the text) in orange.  All of the isochrones assume
[Fe/H] $= -1.33$, $Y = 0.30$, and an age of 11.4 Gyr.  The stars that appear as
purple points in panel (e) are discussed in the text.} 
\label{fig:f8}
\end{figure*}

The models for $Y = 0.25$, when converted from the theoretical to the observed
planes using the {\tt a4xO\_p2} and {\tt a4CNN} BCs, enclose the densest
concentration of stars along the lower RGB in the left-hand panel particularly
well.  However, as in the case if NGC\,6362, our models are unable to explain
the bluest RGB stars (or the reddest stars in the right-hand panel).  That is,
the observed colour variations at a given $M_{F606W}$ magnitude along the RGB
are considerably larger than predicted by our isochrones.
If the bluest giants have high He abundances, $Y$ would have to be much greater
than 0.30 (which was assumed in the isochrone that is represented by long
dashes).  This seems highly improbable.  Not only is a wide range in $Y$ ruled
out by the relatively narrow MS widths that are apparent in Figs.~\ref{fig:f7}a 
and~\ref{fig:f7}b, but the HB of M\,5 is morphologically very similar to the
HB of M\,3, which can be modeled very well without requiring a large He
abundance variation within the cluster (\citealt{dvk17}).

It is worth pointing out that, at the metallicity of M\,5 (and lower [Fe/H]
values; see Fig.~\ref{fig:f10} in Paper I), the $M_{F336W}-M_{F438W}$ colours
are predicted to have very little dependence on $Y$ or the C:N:O abundances in
the vicinity of the TO.  Hence, there is presumably another explanation for the
observed spread in this colour at $M_{F606W} \sim 3.5$--5 (see
Fig.~\ref{fig:f7}a).  Indeed, the same comment can be made to a greater or
lesser extent with regard to the colours that are plotted in the other panels.
Unless the actual abundance variations are much larger than we have
assumed, we suspect that photometric errors are primarily responsible for the
observed dispersion in the TO colours, with perhaps some contributions due to
the presence of binaries and the effects of differential reddening (which is
unlikely to be important in clusters with low $E(B-V)$ values, like M\,5).  As
we have concluded in our analyses of the M\,92 CMDs in \S~\ref{subsec:m92},
photometric scatter would seem to be the only viable explanation for similar
colour spreads at the TOs of the lowest metallicity GCs.

Since chemical differences have much stronger effects on the BCs relevant to
cool giants than to TO stars (see Paper I), the lower RGB stars of M\,5 should
be especially revealing.   This cluster, as in the case of NGC\,6362, seems to
have a bifurcated giant branch on the $(M_{F336W}-M_{F438W})_0,\,M_{F606W}$
diagram; see Figure~\ref{fig:f8}a (the top, left-hand panel), which provides a
magnified view of the SGB and lower RGB of M\,5.  The stars plotted as gray
points can be explained by isochrones that allow for variations in CN strengths
(with perhaps a modest variation in $Y$)--- but not the bluer giants (the black
points), which define a separate sequence of stars.  The fact that the latter
also tend to have redder $(M_{F438W}-M_{F606W})_0$ colours than both the
``normal" giants (see Fig.~\ref{fig:f8}c) and, in particular, the isochrones
in Fig.~\ref{fig:f7}c, rules out the possibility that they have very high He
abundances because, as shown in Fig.~\ref{fig:f7}, stars with higher $Y$ are
predicted to have bluer RGBs on all three CMDs.  Of the several metal abundance
mixtures considered in this investigation, only those that have enhanced C
abundances (see Fig.~\ref{fig:f7} in Paper I) seem to be able to account for
the different behavior of the anomalous giants on the various CMDs.

The effects of high C are illustrated in bottom row of plots in
Fig.~\ref{fig:f8}, which superimpose the lower RGB portions of several
isochrones onto the same CMDs that appear in the upper row.  All of the models
assume $Y = 0.30$, which is probably somewhat higher than the upper limit of the
star-to-star He abundance variation in M\,5.  The dashed, solid, and long-dashed
curves represent, in turn, isochrones for the basic {\tt a4xO\_p2} mixture,
the {\tt a4xC\_p4} mix (i.e., increased C by 0.4 dex), and the {\tt a4xCO}
mixture, in which C has been enhanced by 0.7 dex and O by 0.2 dex.  Note, in
particular, how well these three loci encompass the observed giants in
Fig.~\ref{fig:f8}g.  The dashed curve matches the blue edge of the stars that
have been plotted as gray points, suggesting that they are CN-weak stars with
$Y \approx 0.30$, while the reddest giants are well matched by the models for
[C/Fe] $= +0.7$, [O/Fe] $= 0.6$, and [$m$/Fe] $= 0.4$ for the other $\alpha$
elements.  (The only difference between the {\tt a4xCO} and {\tt a4xO\_p2} 
mixtures is the carbon enhancement of the former.)  The {\tt a4xC\_p4} models,
with [C/Fe] $= +0.4$, lie between those for the {\tt a4xCO} and {\tt a4xO\_p2}
mixtures.  (The inferred {\it variations} in the C abundances should be more 
trustworthy than the absolute abundances implied by the overlays of isochrones
onto observed CMDs given the likelihood that the models are subject to a number
of uncertainties that would mainly affect their $\teff$ and colour zero points.
Another caveat is that the observed colour spread is probably affected to some
extent by photometric errors.)

The C abundance variations that we have considered apparently do not affect 
$M_{F606W}-M_{F814W}$ colours (Fig.~\ref{fig:f8}h), though they may account for
a large fraction of the observed spread in the $M_{F336W}-M_{f606W}$ colours
along the lower RGB at a fixed magnitude (see panel f).  Unfortunately, the
models do not predict sufficiently blue $M_{F336W}-M_{F438}$ colours to match
the stars plotted as black filled circles in panel (e).  However, it is
interesting that the models for the {\tt a4xC\_p4} mix are bluer than those for
the {\tt a4xCO} mixture, despite having a lower C abundance by 0.3 dex.  Since
they also have a lower O abundance by 0.2 dex, one cannot help but wonder if the
bluest giants in Fig.~\ref{fig:f8}e have high C and low O abundances.  (We noted
in Paper I that, although the $F336W$ passband contains an NH band and is
therefore sensitive to the abundance of nitrogen, OH is also quite prominent,
and there are even some CN features in the reddest part of this passband.)  We
did not anticipate this possibility when the project began; consequently, models
for high C, low O mixtures were not generated.  However, we did produce one set
of models, mostly for academic interest at the time, that may have some
relevance for our present discussion.

This set, which will henceforth be referred to as the ``{\tt a4s08C1}" models,
was calculated for the same metal abundances as those listed for the {\tt a4s08} 
mix in Table~\ref{tab:t1}, except that a higher C abundance by 1 dex was
assumed.  In contrast with the enhanced C mixtures discussed so far, such a
large enhancement means that C $>$ O, which has huge implications for low-T
opacities (see \citealt{fd08}) and synthetic spectra (e.g., \citealt[see their
Fig.~2]{vnj17}).  Stars with C $>$ O are predicted to have inflated atmospheres,
lower photospheric pressures, and cooler $\teff$s, especially along the RGB. 
Furthermore, the colours of C-rich stars are very different as a consequence of
replacing oxides by polyatomic molecules involving carbon that produce enormous 
numbers of lines.  Plotted as orange curves in Fig.~\ref{fig:f8} are 11.4 Gyr
isochrones for [Fe/H] $= -1.33$, $Y = 0.30$, and the {\tt a4s08C1} mixture of 
the metals.  These models clearly predict much bluer $M_{F336W}-M_{F438W}$
colours, and much redder $M_{F438W}-M_{F606W}$ colours, than those of the
cluster giants. (The redward offset of the orange curve in panel (h) is due to
the effects of C $>$ O on the model temperatures rather than on the BCs for the
$F606W$ and $F814W$ filters.)  Although such extreme C abundances are not
relevant to M\,5 (or any other GC), there are presumably some combinations
of the C and O (and possibly N) abundances for which isochrones will match the
observed colours of the giants that lie between the orange and blue loci in
Fig.~\ref{fig:f8}e.  Explorations similar to those presented here will need to
be carried out to determine what mixtures of C, N, and O can provide viable
explanations of those stars.  Nevertheless, at this point in time, the
possibility that M\,5 contains a population of C-enhanced stars warrants
serious consideration.
  
\begin{figure}
\includegraphics[width=\columnwidth]{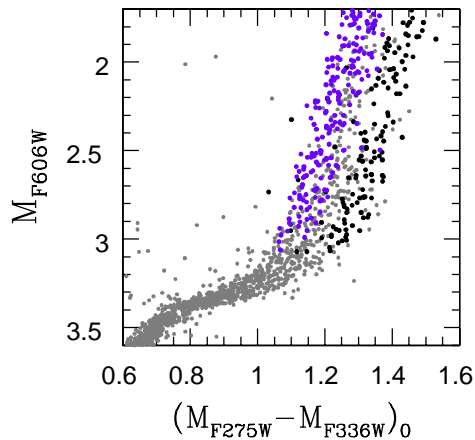}
\caption{As in the previous figure, except that the $(M_{F275W}-M_{F336W})_0$
colours of the lower RGB stars in M\,5 have been plotted.  The stars represented
by black and purple filled circles are the same ones that are similarly
identified in Fig.~\ref{fig:f8}e.}
\label{fig:f9}
\end{figure}

Even though the MARCS spectra were not extended sufficiently far into the UV to
predict BCs for the $F275$ filter, we were curious to know where the putative
C-enhanced stars are located in a CMD in which the $M_{F275W}-M_{F336W}$ colour
is used as the abscissa.  It turns out that, as illustrated by the filled black
circles in Figure~\ref{fig:f9}, they are the reddest stars in such a CMD,
which indicates that they are O-rich.  (Since the $F275W$ passband contains
spectral features due to OH, stars with high O abundances will have stronger
OH, fainter $F275W$ magnitudes, and therefore redder $M_{F275W}-M_{F336W}$
colours.)  Moreover, they are clearly separated from the high N, CN-strong
population that has the reddest $(M_{F336W}-M_{F438W})_0$ colours; note that the
stars plotted as purple filled circles are the same ones that are similarly
identified in Fig.~\ref{fig:f8}e.  These stars apparently have lower O
abundances than those shown as black filled circles, which is consistent with
expectations, since the highest N abundances will be found in a gas that has 
undergone ON-cycling.  Thus, variations in the efficiency of ON-cycling would
provide a natural explanation for the the spread in the $M_{F275W}-M_{F336W}$
colours that is displayed by the purple points in Fig.~\ref{fig:f9}.  As fully
appreciated by \citet{pmb15}, $F275W$ observations clearly provide a very
valuable additional constraint on GC abundances.  Unfortunately, we do not 
have the capability to predict the locations of isochrones for different C, N,
and O abundances in Fig.~\ref{fig:f9} due to the limitations of our current
models.  It would be especially interesting to know where isochrones for [O/Fe]
$= 0.6$ are located in this figure.

Some additional features of Fig.~\ref{fig:f7} are worth pointing out.  In
contrast with NGC\,6362, most of the $\delta\,X$ colour offsets are positive;
i.e., the isochrones had to be shifted to redder colours in order to match the
cluster TOs.  Smaller adjustments would have been found had we used isochrones
for a higher [Fe/H] value, but we found that, in this case, the models do not
fit either the lower RGB or the MS stars nearly as well.  (Note that the
isochrones for [Fe/H] $= -1.33$ and $Y = 0.25$ are almost coincident with the
median MS fiducial sequences from the SGB down to $M_{F606W} \sim 5.5$.)  As
shown below, fits of isochrones to the CMDs of even more metal-deficient GCs
typically require comparable values of $\delta\,X$ in order to match the
observed TOs.

\begin{figure*}
\includegraphics[width=\textwidth]{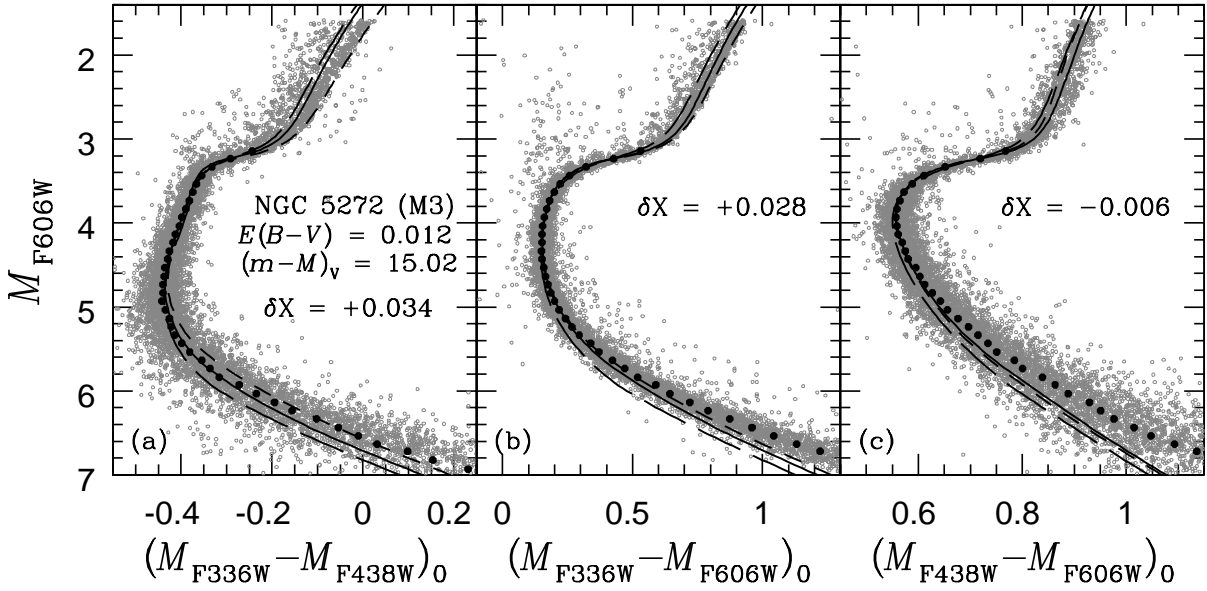}
\caption{Similar to Fig.~\ref{fig:f7}; in this case, WFC3 observations of M\,3
(from NLP18) have been fitted by a 11.9 Gyr isochrones for [Fe/H] $= -1.55$, $Y
= 0.25$, [$\alpha$/Fe] $= 0.4$, and [O/Fe] $= 0.6$.  The long-dashed curve 
assumes $Y = 0.30$.  Although the plot is not included here, no colour offsets
are needed for the isochrones to match the MS and TO colours on the
[($M_{F606W}-M_{F814W})_0,\,M_{F606W}$] diagram, though the models are too red
along the lower RGB by $\sim 0.03$ mag (as found for all other GCs).}
\label{fig:f10}
\end{figure*}

\subsection{NGC\,5272 (M\,3)}
\label{subsec:m3}

The basic properties of M\,3 (NGC\,5272) appear to be quite well determined
as the result of many investigations over the years.  For this system, recent
spectroscopic studies have tended to find [Fe/H] values in the range from
$-1.55$ to $-1.50$ (\citealt{ki03}, \citealt{skg04}, CBG09), with
some preference for the lower, or higher, values if the metallicities are
derived from Fe I, or Fe II, lines, respectively.  M\,3 is known to be nearly
unreddened; e.g., dust maps yield $E(B-V) = 0.011$--0.013 (\citealt{sfd98},
\citealt{sf11}).  The latest simulations of the cluster HB population suggest
that M\,3 has a mean He abundance close to $Y = 0.255$ and a distance
corresponding to $(m-M)_V \approx 15.02$ (\citealt{dvk17}).  Moreover, on the
assumption of very similar cluster properties and chemical abundances, stellar
models are able to explain the periods of member RR Lyrae variables quite
satsifactorily (\citealt{vdc16}).

\begin{figure*}
\includegraphics[width=\textwidth]{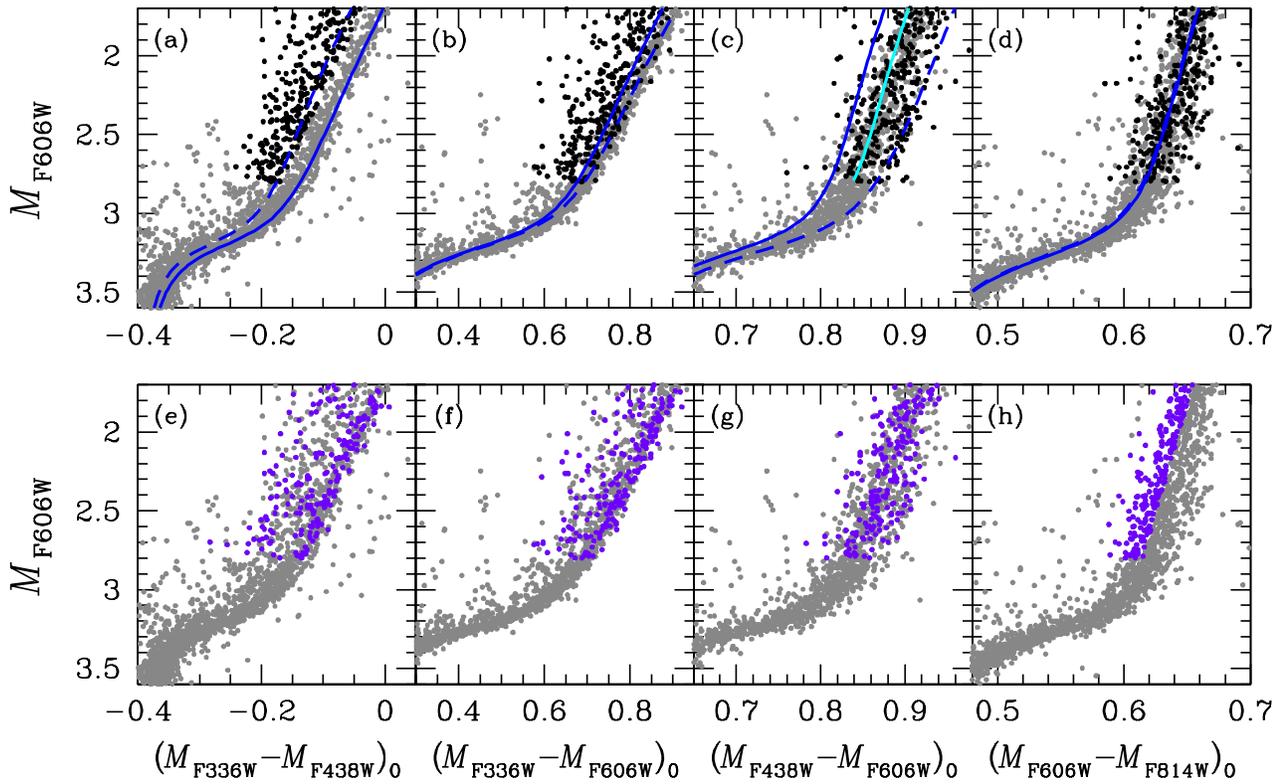}
\caption{Similar to Fig.~\ref{fig:f8}; in this case, expanded views of the
subgiant and lower RGB populations of M\,3 are shown.  {\it Panels (a)--(d)}:
the locations of the stars that were selected in panel (a) are plotted as
black filled circles in all four CMDs.  The nearly vertical line in cyan
represents the lower RGB portion of the same isochrone (for the {\tt a4CNN}
mixture) that appears in panel (c) of the previous figure; this is the reddest
of the three isochrones that are plotted therein.  The isochrones in blue assume
the same age (11.9 Gyr) and [Fe/H] value ($-1.55$), but a higher $Y$ (0.30) and
the {\tt a4ONN} and {\tt a4xCO} mixtures of the metals (solid and dashed loci,
respectively).  {\it Panels (e)--(h)}: the locations of the stars that were
selected in panel (h) are plotted as purple filled circles in all four CMDs.}
\label{fig:f11}
\end{figure*}

If 11.9 Gyr isochrones for [Fe/H] $= -1.55$, $Y = 0.25$, [$\alpha$/Fe] $= 0.4$,
and [O/Fe] $= 0.6$ are fitted to the WFC3 observations of M\,3, assuming $E(B-V)
= 0.012$ and $(m-M)_V = 15.02$, one obtains Figure~\ref{fig:f10}.  As in previous
plots that contain the same three CMDs, the solid and dashed loci were obtained
using the {\tt a4xO\_p2} and {\tt a4CNN} BCs; the long-dashed curve is similar
to the solid curve except that it assumes a higher He abundance ($Y = 0.30$).
The left-hand panel shows that the isochrones enclose the densest concentration 
of lower RGB stars without requiring any additional offset to the predicted
colours other than the $\delta\,X$ value that is needed to fit the TO.
However, as in the case of NGC\,6362 and M\,5, a substantial fraction of
the giants are distributed to much bluer colours than those predicted by the
isochrones at $M_{F606W} \lta 2.8$.  As shown by the long-dashed curve, models
for a helium abundance as high as $Y = 0.30$, which is well outside the range
in $Y$ that has been derived from simulations of the HB population in M\,3
(\citealt{dvk17}), is incapable of explaining the CMD locations of the majority
of the anomalously blue giants. 

Interestingly, the M\,3 giants that lie to the left of the isochrones in
Fig~\ref{fig:f10}a have considerably more overlap with the rest of the lower
RGB stars (those plotted in gray) on the other CMDs than in the case of M\,5
(or NGC\,6362).  This is readily seen by comparing the CMDs in the top row of
Fig.~\ref{fig:f8} for M\,5 with those shown in top row of Figure~\ref{fig:f11},
which similarly plots just the subgiant and and giant-branch stars of M\,3.
Most of the anomalous giants, which are identified by black filled circles, have
redder $(M_{F438W}-M_{F606W})_0$ colours that those predicted by the reddest
isochrone in Fig.~\ref{fig:f10}c; the RGB portion of this isochrone has been
reproduced as the solid curve (in cyan) in Fig.~\ref{fig:f11}c.  Thus, the
majority of the stars that are represented by black points show the photometric
signature of enhanced C abundances in that they have quite red CMD locations
in panel (c) and moderately blue locations in panel (a), which is the expected
consequence of having fainter $M_{F438W}$ magnitudes when CH and CN bands are
stronger.

The TO-to-RGB portions of a few isochrones have been plotted in
Figs.~\ref{fig:f11}a--d.  The solid curve in blue represents an isochrone for
the {\tt a4ONN} mixture to illustrate the predicted colours for a metal
abundance mixture with [N/Fe] $\sim 1.5$ and very low abundances of C and O
([C/Fe] $= -0.8$, [O/Fe] $= -0.4$; see Table~\ref{tab:t1}).  This isochrone
matches the red edge of the CMD that appears in panels (a), which is the
expected consequence of high N and the increased blanketing due to NH in the
$F336W$ passband.  This isochrone also lies close to the blue edge of the
distribution of lower RGB stars in panel (c), which could have been anticipated
because the $F438W$ filter contains spectral features due to CN (which will be
weak) and CH, implying brighter $F438W$ magnitudes and bluer
$(M_{F438W}-M_{F606W})_0$ colours.  (It cannot be concluded from these models
that M\,3 necessarily has stars with such low C abundances because isochrones
for the {\tt a4CNN} mixture, which has [C/Fe] $= -0.3$, provide very similar 
fits to the bluest giants; see Fig.~\ref{fig:f10}c and our results for M\,5 in
Fig.~\ref{fig:f8}g.    Furthermore, the absolute locations of the isochrones
will be affected by whatever errors are present in the BCs and the model
$\teff$s.  The models should be more trustworthy in a relative sense, though
the extent of photometric errors remains a concern.)

Not unexpectedly, C-rich stellar models predict very red
$(M_{F438W}-M_{F606W})_0$ colours.  As shown by the dashed curve in blue,
isochrones for $Y=0.30$ and the {\tt a4xCO} mixture reproduce the location of
reddest giants in panel (c) rather well (just as we found in the case of M\,5;
see Fig.\ref{fig:f8}c).  Of course, the same fit to the observations could be
obtained using stellar models for a somewhat lower He abundance, provided that a
suitably reduced value of [C/Fe] is also assumed, but this could hardly lead to
a reduction in the inferred value of [C/Fe] by more than $\sim 0.2$ dex.
Encouragingly, the solid and dashed loci encompass all of the lower RGB stars of
M\,3 on this particular CMD.  With regard to panels (a) and (b):
although the {\tt a4xCO} models may be relevant to the reddest of the stars
that are plotted as black filled circles, the chemical properties of the bluest
of those stars remain a mystery.  None of our computations for any of the 
mixtures of the metals in Table~\ref{tab:t1} predict such blue colours. 

Recall from Paper I (specifically Fig.~5) that high O by itself has the effect
of producing redder $M_{F336W}-M_{F438W}$ colours (due to the effects of 
enhanced OH).  Reduced O abundances would have the opposite effect, but we
expect that lower O would be accompanied by higher N, which would drive stars
to the red side of the RGB because the BCs for the $F336W$ filter are more
sensitive to NH than to OH.  The same can be said about low C, which is normally
transformed to N via the CN-cycle.  Although the $F336W$ passband contains some
CN features, the BCs for this filter are much more dependent on NH and therefore
on the abundance of nitrogen.  Since high N causes red $M_{F336W}-F{438W}$
colours, only low N remains as a possible explanation of the bluest stars (if
colours are mostly due to variations in the abundances of C, N, and/or O).  It
would not be too surprising, in fact, if this colour is correlated with the
abundance of N from the blue to the red side of the CMD just as the
$M_{F438W}-M_{F606W}$ colour index appears to be directly correlated with C
abundances.)  

\begin{figure}
\includegraphics[width=\columnwidth]{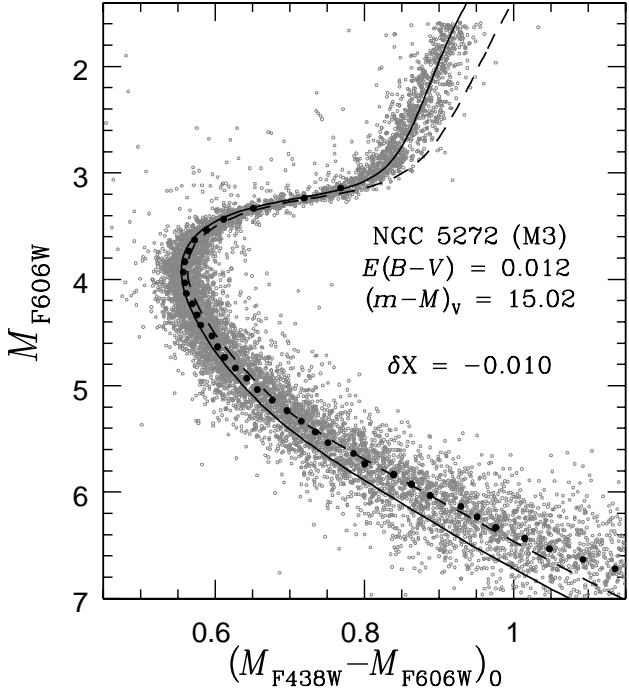}
\caption{Similar to Fig.~\ref{fig:f10}c, except that the location of an 
isochrone for the {\tt a4xCO} mixture (dashed curve) has been fitted to the CMD
of M\,3 along with one for the {\tt a4xO\_p2} mix.}
\label{fig:f12}
\end{figure}

\begin{figure}
\includegraphics[width=\columnwidth]{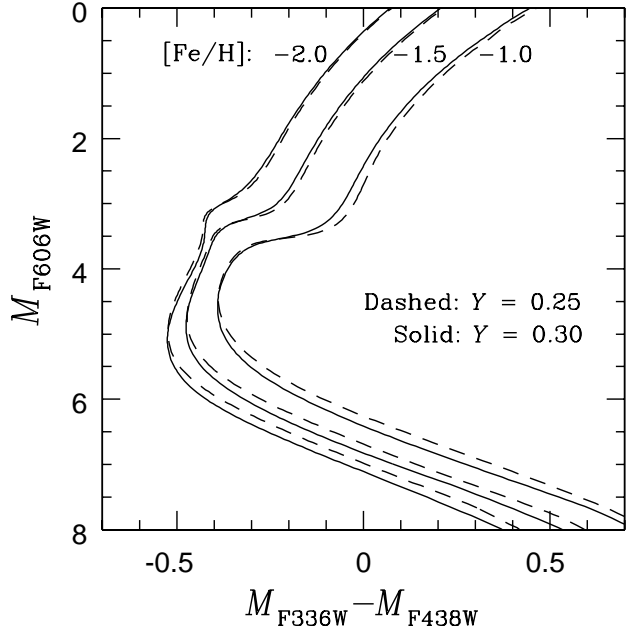}
\caption{Comparisons of 12.5 Gyr isochrones for [$\alpha$/Fe] $=0.4$ and [Fe/H]
$= -2.0$, $-1.5$, and $-1.0$ assuming, for each metallicity, $Y = 0.25$ and
0.30 (as indicated).}
\label{fig:f13}
\end{figure}

The problem is that we do not expect to find stars with very low N abundances
because CN- and ON-cycling always works in the direction of increasing N, and it
has generally been found that, as expected for H-burning reactions, C$+$N$+$O 
$= constant$ to within measuring uncertaintes ($\delta\log$(C$+$M$+$O) $\sim
0.1$--0.15 dex), see, e.g., \citet{ssb96}, \citet{cm05}, \citet{cgl05}.
Nevertheless, observations indicate that there are significant populations of
stars in GCs with $0.0 <$ [N/Fe] $\lta -1.0$, with some indication that the
number of such stars varies inversely with the cluster metallicity (see
\citealt[their Figs.~7, 8, 10, and 11]{cbs05}).  At the present time, large
($\gta 0.4$ dex) star-to-star variations in [CNO/Fe] have been found in only a
few systems that show the strongest evidence for multiple stellar populations
--- notably, NGC\,1851 (\citealt{ygd09}), M\,22 (\citealt{mms12}), and
$\omega\,$Cen (\citealt{mar12}).  It is possible that smaller variations are 
present in other systems, perhaps preferentially in the most massive GCs.
However, it is not known whether stars with very low N abundances have the same
C$+$N$+$O abundance as cluster members with [N/Fe] $\gta 0.0$.  This should be
checked. 

The answer to another question remains elusive: what is the origin of the
relatively large spread in the $(M_{F606W}-M_{F814W})_0$ colour at a fixed
magnitude along the lower RGB?  The $F606W$ and $F814W$ filters are mainly
sensitive to CN (see, e.g., \citealt[their Fig.~4]{ssw11}), but at low 
metallicities and assuming normal abundances of C and N, CN is not sufficiently
important to affect $M_{F606W}-M_{F814W}$ colours by more than $\sim 0.005$ mag
(see the $\delta$(BC) plots shown in Fig.~5 of Paper I).  To be sure, the
effects are larger than this at higher [Fe/H] values; recall Fig.~\ref{fig:f1}c,
which showed that CN-strong giants in 47 Tuc will be $\sim 0.015$ mag bluer than
its CN-weak counterparts at the same $M_{F606W}$.  (Similar results were 
obtained by \citealt{mmr18}.)  The difficulty with CN is
that the effects on BCs will diminish rapidly with decreasing metallicity
because the abundances of both C and N will be reduced at lower [Fe/H] values,
and hence the effect on CN-band strengths will be quadratic.  (For the same
reason $F336W$ magnitudes will be more sensitive to NH, and $F438W$ magnitudes
to CH, than to CN bands, at low metallicities.)  It would therefore seem to be
necessary to have much higher abundances of C and N than in the standard {\tt
a4s21} and {\tt a4CNN} mixtures if CN is responsible for the bluest
$M_{F606W}-M_{M814W}$ colours at the low [Fe/H] values.
 
\begin{figure*}
\includegraphics[width=\textwidth]{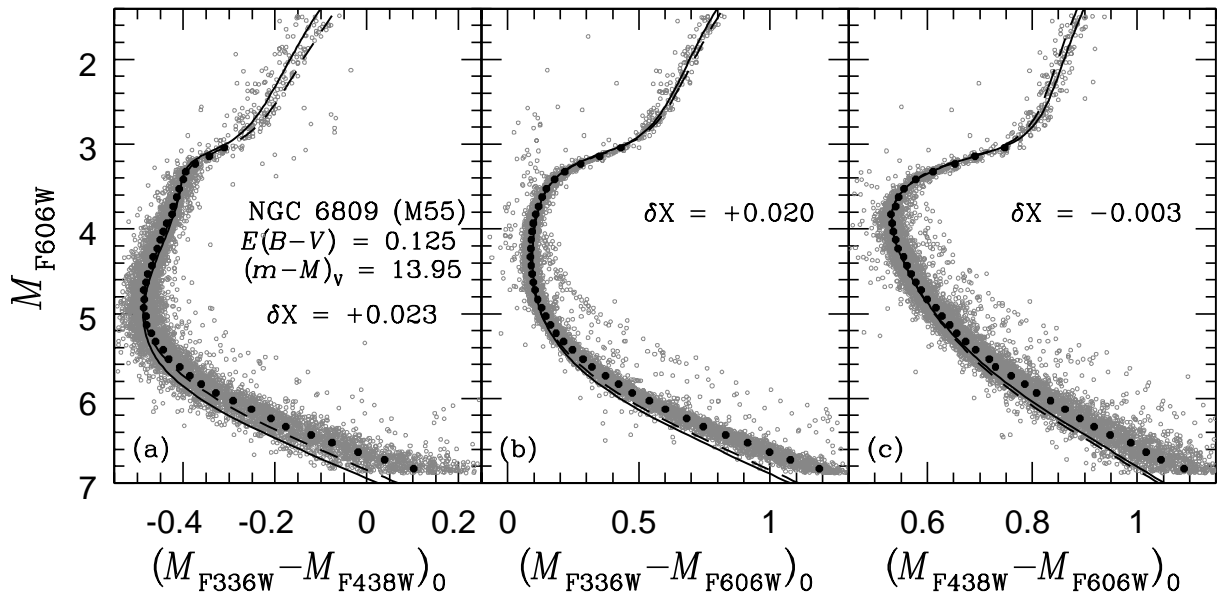}
\caption{Similar to Fig.~\ref{fig:f10}; in this case, WFC3 observations of
M\,55 (from NLP18) have been fitted by 12.8 Gyr isochrones for [Fe/H] $=
-1.93$, $\langle Y_0\rangle = 0.265$, [$\alpha$/Fe] $= 0.4$, and [O/Fe]
$= 0.6$.  Fits to $F606W,\,F814W$ photometry (not shown) are very similar
to those obtained for other GCs; i.e., the isochrones require only a small
offset in the predicted colours ($-0.006$ mag) to match the MS and TO 
observations, but their lower RGB portions are too red by $\sim 0.03$ mag.} 
\label{fig:f14}
\end{figure*}

Helium abundance variations would seem to be the only other way of inducing
large spreads in the $M_{F606W}-M{F814W}$ colour.  An important advantage of
such variations is that their effects on this particular colour index would be
similar for different [Fe/H] values.  Other colours are affected much more by
variations in C, N, and O than He abundance differences.  For instance, as
discussed just above, the reddest giants in M\,3 would appear to have quite high
C, irrespective of whether they have enhanced He abundances.  The bottom row of
panels in Fig.~\ref{fig:f11} shows that the bluest giants in panel (h), where
they are identified by purple points, are are not at all isolated to particular
colour ranges in the CMDs shown in panels (e), (f), and (g), but instead span
the entire colour ranges of the respective CMDs.  In other words, stars with
apparently wide variations in the abundances of C and N, have very similar
$M_{F606W}-M_{V814W}$ colours.  According to, e.g., \citet[their Figs.~10 and
11]{cbs05}, the star-to-star variations of [C/Fe] and [N/Fe] within GCs range
from 0.8 to 1.3 dex and 1.7 to 2.8 dex, respectively.  Note, as well, that
small fractions of the stars in some of the clusters that they considered appear
to have [C/Fe] $\sim 0.2$--0.3 and [N/Fe] $\gta 1.7$\footnote{These results
assume solar abundances that are about 0.2 dex higher than those determined
subsequently by \citet{ags09}; consequently, the reported values of [C/Fe] and
[N/Fe] should be increased by 0.2 dex to put them on the Asplund et al.~scale.}
(also see \citealt{bhs04}).  It is tempting to conclude that helium is primarily
responsible for the variation in the observed $M_{F606W}-M_{F814W}$ colour
along the lower RGB, and that CN plays a secondary role.

The MS of M\,3 also seems to be somewhat anomalous in that its slope in
Fig.~\ref{fig:f10}c is much shallower than that predicted by our isochrones
for the {\tt a4xO\_p2} mixture.  Such a strong deviation between the models and
the observations, which begins at $M_{F606W} \approx 4.5$, was not seen in the
GC CMDs considered thusfar.  Furthermore, the spread in the
$(M_{F438W}-M_{F606W})_0$ colours at a given magnitude seems unusually large,
which gives one the distinct impression that $F438W$ magnitudes are more
problematic than either those for the $F336W$ or $F606W$ passbands.  While it
would not surprising that there would be systematic errors in any of the colours
with increasing $F606W$ magnitude along the MS (due, e.g., to problems with the
predicted temperatures of stellar models or systematic errors in the BCs that
are correlated with the temperatures or colours of stars), the large
morphological differences may be another indication that some stars in M\,3
have unusually high C abundances.  In fact, the MS slope on the
$(M_{F438W}-M_{F606W})_0\,M_{F606W}$ CMD is predicted to be a function of the
C abundance.  This is illustrated in Figure~\ref{fig:f12}, which shows that
isochrones for the {\tt a4xCO} mixture provide a much better fit to the MS
fiducial of M\,3 than the {\tt a4xO\_p2} models.

We searched for other explanations, but unsuccessfully.  For instance, we
generated grids of evolutionary tracks and isochrones for alternative values
of [Fe/H] (between $-1.6$ and $-1.45$), and for reduced O abundances by 0.2
dex, without finding a significant change to the predicted slope of the MS.
(Higher $Y$ exacerbates the problem.)  We also ruled out the possibility that
interpolation errors in the relevant $\delta$(BC)--$\log\,g$ relations (like
those plotted in Fig.~\ref{fig:f2} of Paper I) are responsible for the shallow
MS slope, but essentially the same slope is predicted by the CV14
transformations.  It would therefore appear that enhanced C abundances are
favoured by both the observed MS slopes and the fits of isochrones to GC giants
with the reddest $M_{F438W}-M_{F606W}$ colours; i.e., the assumption that GC
stars have [C/Fe] $\lta 0.0$ may not be correct.  Although such results
as those shown in Fig.~\ref{fig:f10} seem quite agreeable, at least in an 
overall sense, we suspect that it could be quite fruitful to explore the
consequences of wider variations in the abundances of C and N than in our
current models.

As regards the small discrepancies at $M_{F606W} \sim 5.4$ in Fig.~\ref{fig:f10}a,
they can be reduced by adopting a lower metallicity or a higher He abundance.
Consider Figure~\ref{fig:f13}, which shows that the magnitude difference between
the roughly horizontal transition from the TO to the RGB at $M_{F606W} \lta
3.5$ and the MS at $M_{F606W} \gta 5.8$ varies inversely with [Fe/H] (and age)
and directly with $Y$.  The morphology of an isochrone in the vicinity of the
TO is also quite a strong function of the metallicity --- much more so than in
optical CMDs.  Indeed, the location and shape of the principal photometric
sequence of M\,3 on the [$m_{F336W}-m_{F438W},\,m_{F606W}$]-diagam provides a
compelling argument that this cluster has a metallicity within $\pm 0.1$\ dex
of [Fe/H] $= -1.55$.  Curiously, isochrones for the same [Fe/H] value and age
are predicted to be slightly redder at the TO ($M_{F606W} \sim 4.0$) if they
assume a higher $Y$, which is contrary to expectations given that the turnoff
$\teff$\ is predicted to be somewhat hotter (by 20--40 K) if $Y = 0.30$ than if
$Y = 0.25$.  The same behavior is, however, found if the CV14 transformations
are used to transpose the models from the theoretical to the observed plane.

\begin{figure}
\includegraphics[width=\columnwidth]{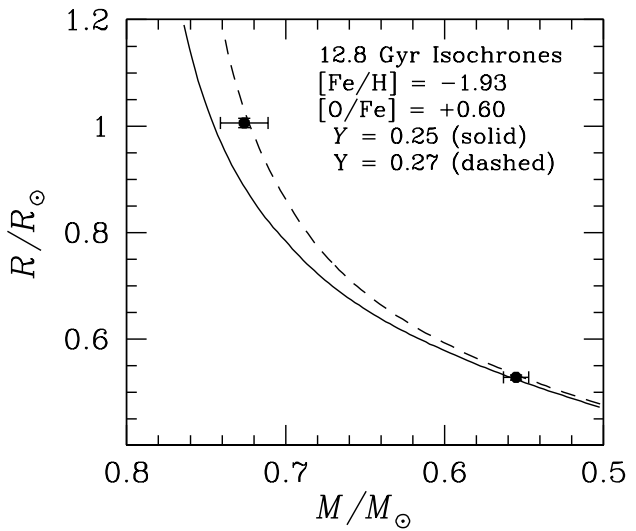}
\caption{Mass-radius relations from 12.8 Gyr isochrones for [Fe/H] $= -1.93$
and the indicated values of [O/Fe] and $Y$ are compared with the properties of
the binary V54 (filled circles and error bars) that were derived by
\citet{ktd14}.}
\label{fig:f15}
\end{figure}

The $\delta\,X$ colour offsets in Fig.~\ref{fig:f10} are somewhat smaller than
those determined for M\,5, but in the same sense.  As in the latter cluster,
they suggest that the predicted $M_{F336W}$ magnitudes near the TO are slightly
too bright.  These offsets would be reduced if a higher metallicity were
adopted, but doing so would increase the discrepancy between the predicted and
observed $M_{F438W}-M_{F606W}$ colours.  Moreover, models for a higher [Fe/H]
value would not be able to match the CMD locations of either the lower RGB stars
or the MS stars at $M_{F606W} \sim 5.5$ quite as well.  If anything, these
observations suggest that M\,3 may have a slightly lower metallicity (closer
to [Fe/H] $= -1.6$).

\subsection{NGC\,6809 (M\,55)}
\label{subsec:m55}

In their recent study of M\,55, \citet{vd18} concluded from their analysis
of its CMD, the cluster RR Lyrae variables, and the eclipsing binary V54
(\citealt{ktd14}) that it has a reddening in the range $0.12 \lta E(B-V) \lta
0.13$, as found from dust maps (\citealt{sfd98}, \citealt{sf11}), $(m-M)_V =
13.95 \pm 0.05$, which was also shown to agree very well with the distance
derived from MS fits to local subdwarfs, [Fe/H] $= -1.85 \pm 0.1$, in agreement
with spectroscopic findings (e.g., CBG09), and [O/Fe] $= 0.5 \pm 0.1$.
Furthermore, the simulations of the cluster HB populations that were presented
in the same investigation indicated that most of the core He-burning stars have
$0.25 \lta Y_0 \lta 0.27$ with only $\sim 10$\% of them having slightly higher
He abundances. In keeping with our analyses of M\,3 and M\,5, we have
adopted the metallicity given by Carretta et al.~($-1.93$ to be specific),
[O/Fe] $= 0.6$ (with [$m$/Fe] $= 0.4$ for the other $\alpha$ elements), and the
aforementioned cluster parameters.  Under these assumptions, 12.8 Gyr isochrones
for approximately the mean He abundance, $\langle Y_0\rangle = 0.265$, provides
reasonably good fits to the WFC3 CMDs for M\,55 --- as shown in
Figure~\ref{fig:f14} --- when they are transformed to the observational planes
using the {\tt a4xO} and {\tt a4CNN} BCs.

\begin{figure*}
\includegraphics[width=\textwidth]{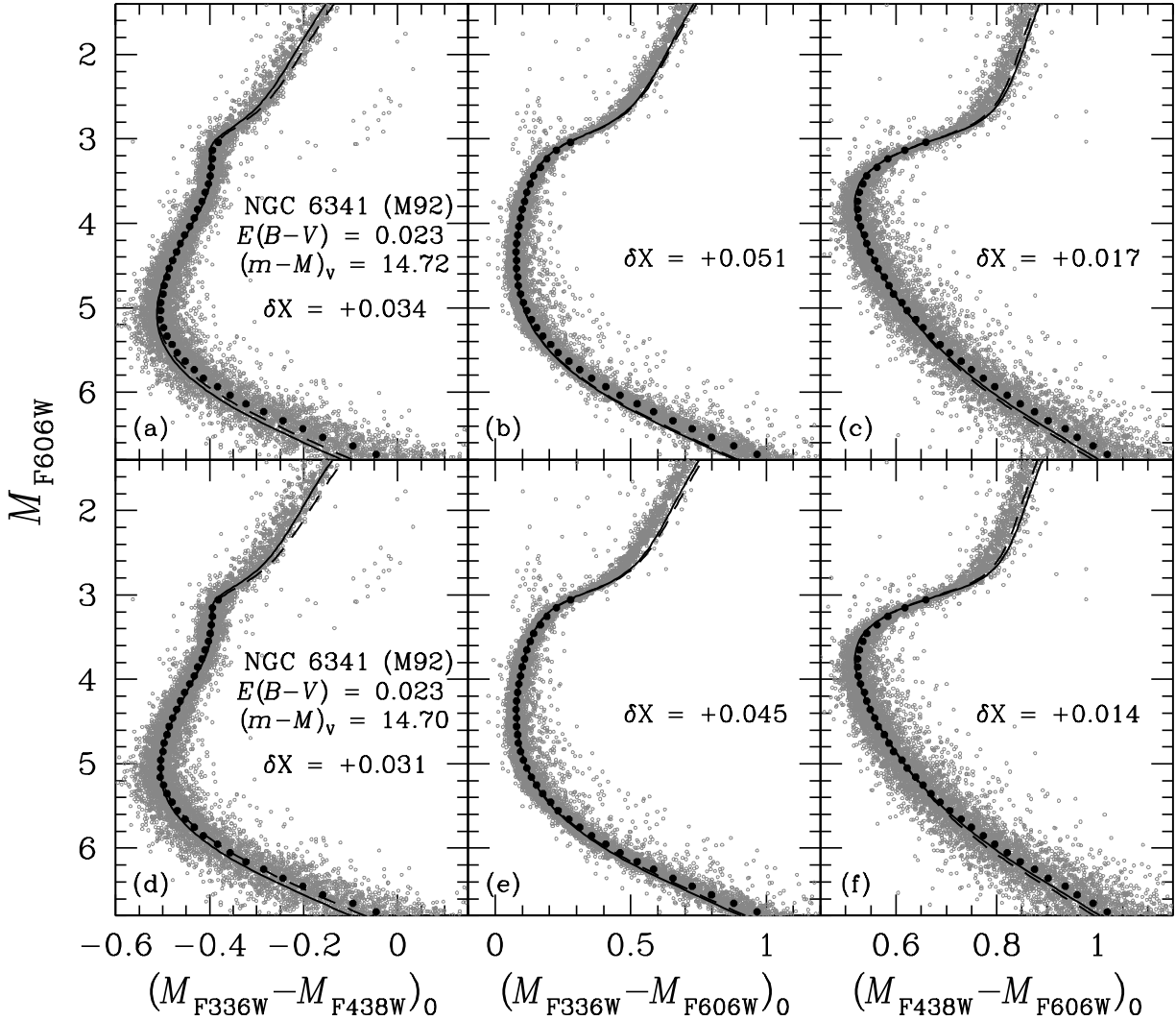}
\caption{Similar to Fig.~\ref{fig:f10}; in this case, panels (a) to (e) show
fits of 12.6 Gyr isochrones for [Fe/H] $= -2.35$, $\langle Y_0\rangle = 0.265$,
[$\alpha$/Fe] $= 0.4$, and [O/Fe] $= 0.6$ to WFC3 observations of M\,92 (from
NLLP18). Panels (d) to (f) show similar fits, but using 12.3 Gyr isochrones for
[Fe/H] $= -2.20$.  In either of these cases, the isochrones provide very good 
fits to $F606W,\,F814W$ observations (not shown) along the upper MS and TO if
they are adjusted by $\delta$\,X $\approx -0.004$, though the predicted
giant-branch locations are too red by $\sim 0.04$ mag.  As already mentioned,
such fits are common to all of the GCs that we have considered.} 
\label{fig:f16}
\end{figure*}

The left-hand panel, in particular, provides strong support for the adopted
[Fe/H] value given that the morphology of the
[$(M_{F336W}-M_{F438W})_0,\,M_{F606W}$]-diagram depends quite sensitively on
metallicity (recall our discussion of Fig.~\ref{fig:f13}).  The separation of
the solid and dashed loci along the lower RGB is also quite a strong inverse
function of [Fe/H]; compare these predictions with those applicable to M\,3
in Fig.~\ref{fig:f10} and NGC\,6362 in Fig.~\ref{fig:f5}.  As for all of the
other GCs considered in this study, none of our stellar models is able to
reproduce the CMD locations of the bluest, lower RGB stars, though we suspect
that at least some of the giants may have relatively low N abundances.  On the
other hand, at very low metallicities in particular, colours are only weakly 
dependent on the abundances of the light elements and it seems doubtful that
stellar models for reduced N by 0.5--0.7 dex or so would predict the relatively
large variation in the $(M_{F336W}-M_{F438W})_0$ colours at a given $M_{F606W}$
to the left of the solid curve in Fig.~\ref{fig:f14}a, which assumes [N/Fe]
$= 0.0$.  Perhaps something other than metal abundance variations is
responsible for the relatively large width of the lower RGB of M\,55 in this
panel.

Aside from this difficulty and
the usual deviations between the predicted and observed MS slopes, the
models provide satisfactory fits to the WFC3 photometry.  Note that the
$\delta\,X$ offsets are slightly less than, though similar to, those found
for M\,3.  Finally, given the importance of the binary constraint, we show in
Figure~\ref{fig:f15} that the mass-radius relations from the isochrones for
the same metal abundances, but for He abundances that have been inferred from
HB simulations (\citealt{vd18}), are consistent with the masses and radii that
have been determined for the eclipsing binary member, V54, by \citet{ktd14}.  

\subsection{NGC\,6341 (M\,92)}
\label{subsec:m92}

M\,92 is a suitable representative of the most metal-deficient GCs in the
Milky Way given that it is subject to relatively low reddening ($E(B-V) = 
0.019$--0.023 according to the dust maps of \citealt{sfd98} and \citealt{sf11}),
most determinations of its apparent distance modulus lie in the range $14.65 \le
(m-M)_V \le 14.79$ (see the summary of published results given by \citealt[their
Table 1]{vd18}), and spectroscopic studies generally find [Fe/H] $\sim -2.35$
(\citealt{ki03}, CBG09).  However, a significantly lower metallicity,
[Fe/H] $\lta -2.6$, has been recently derived by \citet{rs11};
consequently, the metal abundance of M\,92 may be the least well determined
of its basic properties.  Nevertheless, if we adopt [Fe/H] $= -2.35$ (from
CBG09) and the same values of $\langle Y_0\rangle$, [$\alpha$/Fe], and
[O/Fe] that were adopted for M\,55, which has a very similar HB morphology,
we obtain the fits of 12.6 Gyr isochrones to the WFC3 photometry of M\,92 that
are shown in panels (a) to (c) of Figure~\ref{fig:f16}.  These results assume
$E(B-V) = 0.023$ and $(m-M)_V = 14.72$, which are supported by the fits of ZAHB
models to the lower boundary of the distribution of cluster HB stars (see, e.g.,
VBLC13).

\begin{figure*}
\includegraphics[width=\textwidth]{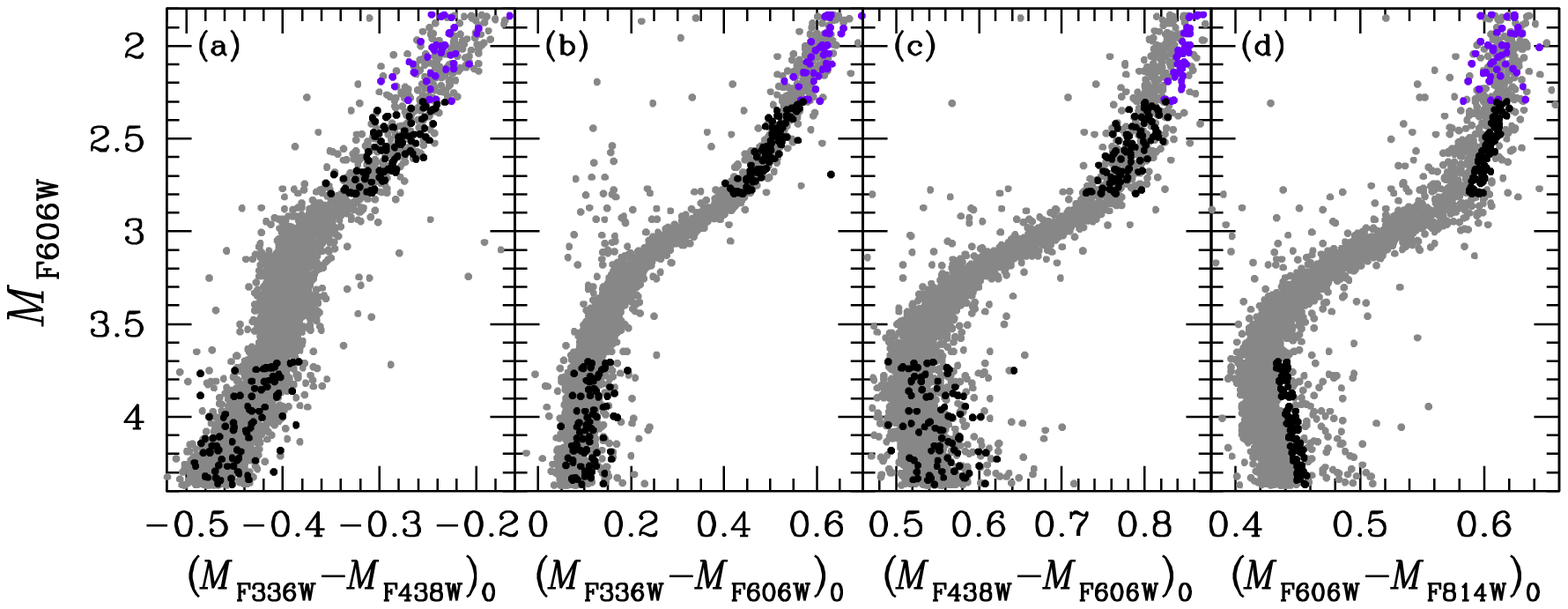}
\caption{Magnified view of the upper MS, TO, and lower RGB portions of the
M\,92 CMDs.  Samples of the giants and upper MS stars that lie within the 
narrow bands in panel (d), which have the reddest $(M_{F606W}-M_{F814W})_0$
colours, are plotted in all four panels as black filled circles.  Similarly, a
selection of the giants in panel (c) with the reddest $(M_{F438W}-M_{F606W})_0$
colours appear in all four panels as purple filled circles.} 
\label{fig:f17}
\end{figure*}

Although the distinctive hook feature at $M_{F606W} \sim 3.1$ in the top,
left-hand panel is reproduced very well by the models, it is somewhat 
disconcerting that rather large $\delta\,X$ colour offsets are needed in
Figs.~\ref{fig:f16}a and~\ref{fig:f16}b in order to register the isochrones to
the observed TOs.  The fact that the isochrones need to be shifted to redder
colours possibly suggests that the adopted metallicity is too low, which is
also indicated by the differences between the isochrones and the median fiducial
sequence at $M_{F606W} \gta 5.2$ in the left-hand panel.  Not only would a
lower metallicity increase these discrepancies along the lower MS, but larger
colour offsets (though only by small amounts) would also be needed in order for
the isochrones to match the TO and MS observations because isochrones for lower
[Fe/H] values are bluer.
%

On the other hand, if the metallicity given by CBG09 is increased by
0.15 dex to [Fe/H] $= -2.2$, one obtains remarkably good fits to the MS and
TO observations of M\,92; see Figs~\ref{fig:f16}d--f.  These
results appear to preclude the very low metallicity that was derived by
\citet{rs11}.  Apparently, the $\delta\,X$ offsets are not very dependent on
the assumed [Fe/H] value, as they are still quite large even if M\,92 has
[Fe/H] $= -2.2$.  At such low [Fe/H] values, variations in the mixture of the
metals should have almost no effects on the predicted colours (see Paper I), so
the large colour offsets may be due, at least in part, to problems with the BCs
or the model $\teff$\ scale. 
Alternatively, M\,92 may have a higher reddening than we have assumed, as
suggested by \citet{ksb98}.  Since $E(F336W-F606W) = 2.26\,E(B-V)$ (see CV14),
an increased reddening by only 0.01 mag would reduce the discrepancy between
the predicted and observed $(M_{F336W} - M_{F606W})_0$ colours by 0.026 mag
(i.e., by slightly more than a factor of two).

Perhaps the main difficulty with Figs.~\ref{fig:f16} is that the models are
unusually red along the giant branch.  Unlike M\,55, M\,3, and M\,5, there are
very few stars with redder $(M_{F438W}-M_{F606W})_0$ colours than the
isochrones (see panels c and f).  This could be suggesting that M\,92 stars
have lower C abundances than those residing in the other three GCs, but with
so many uncertain factors at play in the fits of isochrones to observed CMDs,
the correct explanation could easily be something else.  In fact, better
consistency in this regard would be obtained if M\,92 has [Fe/H] $\lta -2.5$
(as derived by \citealt{rs11}), since the lower RGB portions of isochrones for
such low metallicities are significantly bluer than those shown in
Fig.~\ref{fig:f16}.  However, in this case, the models would not match the MS
fiducial at $M_{F606W} \gta 5.2$ nearly as well --- though the discrepancies
could still be within the uncertainties associated with the assumed distance,
the synthetic BCs, and the predicted temperatures.  

More than any of the other GCs that we have considered, M\,92 raises the 
concern that photometric errors may be playing a significant role in the 
observed colour spreads at a given evolutionary stage.  According to Paper I,
none of the metal abundance variations that we have considered in this project
should have any effects on the magnitudes and colours of TO
stars at [Fe/H] $\lta 2.0$, and yet the widths of the various CMDs at the TO
are $\gta 0.05$ mag.  The only possible ``chemical" explanation would seem to
be He abundance variations, but our isochrones indicate that the effects of
$\delta\,Y = 0.04$, which is probably close to the maximum such variation in
M\,92, would affect TO colours by no more than 0.01 mag.  This leaves
photometric scatter as the most likely explanation.

This suggestion is supported by Figure~\ref{fig:f17}.  If the M\,92 stars with
the reddest $(M_{F606W}-M_{F814W})_0$ colours, at a given $M_{F606W}$, have
somewhat lower He abundances than those with bluer colours, such stars should
also have the reddest $(M_{F438W}-M_{F606W})_0$ colours (assuming that the
colours are not affected by any metal abundance variations that might be
present, as predicted by our stellar models for very low metallicities).
However, the reddest stars in panel (d), specifically those plotted as black
filled circles, which have been constrained to lie within narrow bands,
have very wide colour distributions in the other panels.  One has the visual
impression that the MS disributions in panels (b) and (c) are skewed just
slightly to the red, but not by very much.  Similarly, if a selection is made
in panel (c) of the lower RGB stars with the reddest $(M_{F438W}-M_{F606W})_0$
colours (e.g., the giants identified by filled circles in purple), they have
very broad colour distributions in all of the other CMDs.  In fact, stars that
have very similar colours in any one of the four CMDs span wide colour ranges
in the other three CMDs.  The thickness of the principal photometric sequence
in M\,92 is apparently due mostly to photometric errors.
 
This does not call into question our findings in the case of M\,5 and M\,3;
specifically the identification of C-rich, N-rich, and O-rich populations in
those GCs, because their locations in the various CMDs differ in the expected
ways due to the effects of CH and CN on $F438W$ magnitudes, of NH on $F336W$
magnitudes, and of OH on $F275W$ magnitudes.  (The absolute abundances of C,
N, and O are also much higher in these clusters than in M\,92, and consequently,
spectral features due to molecules that involve atoms of these elements will be
much stronger.  To a considerable extent, our isochrones are able to explain
the observed CMDs.)  However, photometric errors may explain the unusually blue
colours of some of the cluster giants, since the scatter due to, e.g., the
blending of images of pairs or groups of stars will be preferentially to the
blue side of the RGB (\citealt{bs09}).  Complementary spectroscopic studies
of samples of the bluest, and the reddest, lower RGB stars with the very best
photometry would undoubtedly be very fruitful.

\section{Summary and Discussion}
\label{sec:sum}

This investigation has shown that Victoria-Regina isochrones together with the
BCs that have been derived from the latest MARCS model atmospheres and synthetic
spectra are able to reproduce the morphologies of the CMDs that can be
generated from {\it HST} UV Legacy Survey WFC3 photometry (\citealt{pmb15},
NLP18) surprisingly  well.  Even in an absolute sense,
the models appear to be able to match observed UV, optical, and near-IR
magnitudes and colours to within $\sim 0.03$ mag, which is easily within the
total uncertainty due to possible errors in the $\teff$ scale, the zero points
of both the synthetic and observed photometry, the adopted cluster properties
(i.e., distances, reddenings, and metallicities), and remaining deficiencies of
the synthetic spectra.  Of particular note is the capability of the isochrones
to reproduce the development of the ``kink" that appears near the TO in
$(M_{F336W}-M_{F438W})_0,\,M_{F606W}$ CMDs at [Fe/H] $\sim -1.8$, becoming
quite a pronounced feature as the metallicity decreases below [Fe/H] $\sim
-2.0$.  There is some tendency for the predicted colours to deviate to the blue
of observed colors along the LMS, which may be telling us that the assumed
carbon abundances are too low, but there are no obvious deficiencies in the
fitting of upper MS, TO, and the densest concentrations of lower RGB stars.  As
shown in Paper I, the improved BCs, especially at UV wavelengths, are superior 
to those provided by CV14, which are based on the previous generation of MARCS
models (\citealt{gee08}), and apparently those derived from Kurucz model
atmospheres and synthetic spectra as well (see \citealt{mcc17}).

The main thrust of this project has been to examine the consequences of
variations in the abundances of C, N, and O for GC CMDs.  Our focus has
been on lower RGB stars where the effects of light-element abundance variations
are predicted to be quite substantial (see Fig.~5 in Paper I).  The giant-branch
populations of 47 Tuc, NGC\,6362, M\,5, M\,3, M\,55,
and M\,92, which span the range in [Fe/H] from $\sim -0.7$ to
$\sim -2.3$, show qualitatively similar morphologies, but with colour spreads
that vary with metallicity.  The most straightforward CMD to explain seems to
be the $(M_{F438W}-M_{F606W})_0,\,M_{F606W}$ diagram, as the star-to-star colour
variation at a given magnitude is predicted to be strongly correlated with the 
abundance of carbon (in the sense that the C abundance increases in the 
direction from blue to red).  The dependence is unlikely to be strictly linear,
however, given the existence of He abundance variations, but carbon has a much
greater effect on the $(M_{F438W}-M_{F606W})_0$ colour than He. 

The isochrones that have been fitted to $F438W,\,F606W$ observations of M\,5
and M\,3 suggest that some of their stars may have [C/Fe] as high as $\sim +0.5$
dex.  It is difficult to assess the reliability of this prediction because
the $\teff$s and colours of stellar models depend on so many factors, each with
their own uncertainties, but we note that isochrones for the low C abundances
that result from CN- or ON-cycling provide good fits to the blue edges of
the colour distributions and that rather high C is needed to obtain comparable
fits to the reddest stars. Thus, the observed colour spreads suggest quite a
wide range in abundance of carbon.  By comparison, field halo stars with high
values of [$\alpha$/Fe] have [C/Fe] $\sim 0.2$--0.3 (\citealt{ncc14}), which is 
similar to the C abundances that have been derived for GCs by \citet{bhs04} and
\citet{cbs05} once their determinations have been adjusted to be on the
\citet{ags09} solar scale.  It is worth mentioning that the eclipsing binary
stars in GCs favor stellar models that assume [O/Fe] $\gta 0.6$ and [$m$/Fe]
$\approx 0.4$ for the other $\alpha$ elements.  If [O/Fe] $\lta 0.6$ in GCs,
as found in solar neighborhood Pop.~II stars (\citealt{fna09}, \citealt{ral13},
\citealt{ncc14}, \citealt{ans19}), then an enhancement in the
abundance of carbon would help to satisfy the binary constraint (since the
observed $M$--$R$ relations for GC binaries seem to require high values of
[CNO/Fe]).

The $(M_{F336W}-M_{F438W})_0,\,M_{F606W}$ diagram can be used to identify stars
with different N abundances (\citealt{pmb15}), and/or C abundance variations
since increased C will result in fainter $M_{F438W}$ magnitudes,
and therefore bluer $(M_{F336W}-M_{F438W})_0$ colours.  In the case of GCs, this
CMD typically consists of two components --- a dense concentration of giants
with red colours, and a bluer, usually somewhat more diffuse population that
appears to be a separate sequence in clusters with intermediate metallicities
(such as NGC\,6362 and M\,5) or one that overlaps with the reddest stars in
lower metallicity systems (e.g., M\,3, M\,55, M\,92).  Our isochrones generally
provide good fits to the first component, but not the second, if they assume 
[CNO/Fe] $\approx 0.44$ and allow for variations in the abundances of C and N
that would be produced by CN-cycling.  In particular, models for [C/Fe] $=$
[N/Fe] $= 0.0$, which are relevant to ``CN-weak" stars, and those for [C/Fe]
$= -0.3$ and [N/Fe] $= 1.13$, which is representative of the abundances in
``CN-strong" stars, contain the observed spreads in the $(M_{F336W}-M_{F438W})_0$
colours of the redder giants, as well as the metallicity dependence of these
spreads, remarkably well.

However, our current models are unable to explain the $(M_{F336W}- M_{F438W})_0$
colours of the bluer RGB component if they have [N/Fe] $\ge 0.0$.  As noted in
the previous paragraph, isochrones for [N/Fe] $= 0.0$ coincide with the blue
edges of the red RGB populations in GCs, from which one might conclude that the
bluest giants must have much lower N abundances.  In fact, this would seem to be
the only possible ``chemical" solution to the puzzle.  Although some of our
metal abundance mixtures allow for low or high C abundances, and others allow
for low or high O abundances, the isochrones for such mixtures are significantly
too red (see Figs.~\ref{fig:f8} and~\ref{fig:f11}).  Thus, the very blue
$(M_{F336W}-M_{F438W})_0$ colours cannot be attributed solely to the effects of
C and/or O abundance variations on the BCs for the $F336W$ and/or the $F438W$
filters.  Whether or not weak NH, due to very low N abundances, can explain the
bluest $(M_{F336W}- M_{F438W})_0$ colours remains to be determined as we did 
not compute any models for low values of [N/Fe].  However, it is a concern that,
at the lowest metallicities, variations in C, N, and O are predicted to have
very small effects on colours, even along the lower RGB.  It may well turn out
that low N is not the answer.  Indeed, an explanation of the anomalously blue
giants in terms of binaries and evolved blue stragglers may be an especially
promising alternative possibility; see \citet{mms19}.

Regardless, it would still be of considerable interest if GCs contain populations
of stars with very low N abundances.  Even though it is often assumed that GC
stars have [N/Fe] $= 0.0$ or higher, there have been a number of investigations
of metal-deficient field dwarfs over the years that have reported values of
[N/Fe] between $\sim -0.2$ and $\sim -0.7$ (e.g., \citealt{tl84}, \citealt{la85},
\citealt{cbk87}).  Importantly, such stars appear to be present in GCs as well
(see \citealt[their Figs.~6, 8, and 10]{cbs05}).  On the other hand, the
reliability of such findings is questionable given that ``nitrogen abundances
are notoriously difficult to determine with accuracy" (\citealt[p.~57]{kra94}).
Some of the causes of this uncertainty are described by \citet{scp05}, who
found, for instance, that there are systematic differences amounting to 0.4 dex
between the N abundancs that are derived from NH and CN bands.  Nevertheless,
Spite et al.~concluded that the low-luminosity, unmixed giants in their sample
of extremely metal-poor stars ([Fe/H] $< -2.7$) have N abundances that extend
as low as [N/Fe] $\sim -1.0$.  They note that such abundances could have been
produced by the same Type II supernovae that are believed to be responsible for
releasing so much oxygen into the early universe, as suggested by the models of
\citet{mm02}, for instance.

In this regard, it seems pertinent to recall that our consideration of the
$(M_{F275W}-M_{F336W})_0$ colours of the the anomalous population of giants
revealed that these stars are O-rich, which probably means that they have
[O/Fe] values that are close to the maximum value (i.e., [O/Fe] $= 0.6$)
since their UV colours overlap with those of normal giants (see
Fig.~\ref{fig:f9}).  Unfortunately, it is not possible at this time to
investigate the dependence of $(M_{F275W}-M_{F336W})_0$ colours on the O
abundance, as we do not have the capability to predict BCs for the $F275W$
filter.  Still, it is an intriguing possibiity that the stars with apparently
very low N and high O abundances may have formed out of gas that was not
``contaminated" by the chemical evolution that occurred during the formation of
GCs (which produced the observed C--N--O--Na--Mg--Al--Si correlations and
anticorrelations).

The $(M_{F606W}-M_{F814W})_0,\,M_{F606W}$ diagram remains something of a
mystery as our models fail to explain the observed colour spreads along the
lower RGB at a given magnitude.  Since the $F606W$ and $F814W$ passbands are
mostly sensitive to CN (see \citealt{ssw11}), our failure could simply be an
indication that we have not considered sufficiently high N abundances
and/or the optimum ratios of C:N to maximize predicted CN strengths.  This needs
to be investigated.  However, we suspect that this may not be a viable solution
at the lowest metallicities given the decrease in the abundances of both C and
N with decreasing [Fe/H] and the concomitant rapid decrease in CN strengths.
Allowing for larger He abundance variations may help to resolve this problem,
as would higher [CNO/Fe], which would tend to reduce the separation in colour
between the turnoff and the lower RGB.  We doubt that the discrepancies between
the predicted and observed colours are due mostly to errors in the model $\teff$
scale because, for the most part, our isochrones provide reasonable fits to the
other CMDs that we have considered without having to apply temperature
corrections.

Knowing the total C$+$N$+$O abundance in GCs is exceedingly important for fits
of isochrones to observed CMDs (notably to the difference in colour between the 
TO and lower RGB) and for their ages.  The evidence from binary stars seems
compelling that [CNO/Fe] $\gta 0.44$, which is obtained for a
primordial mixture with [C/Fe] $=$ [N/Fe] $= 0.0$ and [O/Fe] $= 0.6$.
Stellar models that assume [O/Fe] $=$ [$\alpha$/Fe] $= 0.4$ are precluded, not
only by the binaries but also by the the high N abundances that are typically
derived for cluster giants.  For instance, if nearly all of the C and O in the
high oxygen primordial mixture were converted to N via the CNO-cycle, the
resultant N abundance would be close to [N/Fe] $= 1.5$, which is generally
found in the majority of GCs (see, e.g., \citealt{cbs02}, \citealt{bcs04},
\citealt{sbh05}).  Note that if the maximum values of [N/Fe] were closer to
$+1.7$ (for which there is some spectroscopic support, see \citealt{bhs04},
\citealt{cbs05}), such high abundances by themselves (i.e., without any
contribution from C and O) would imply [CNO/Fe] $= 0.65$, which is 0.2 dex
higher than we have assumed in the majority of our computations.  Any increase
in [CNO/Fe] would necessarily result in reduced ages at a given TO luminosity;
consequently, the ages of $\lta 12.8$ Gyr that we have obtained in this study
using isochrones for [CNO/Fe] $= 0.45$, on the assumption of well-supported
distances and reddenings, may be upper limits to their actual ages.

Although our stellar models appear to be able to reproduce the observed colours
of GC stars and the widths of photometric sequences in cluster CMDs quite
well, that success can be claimed only if the inferred abundances of C, N, and
O from the superposition of the isochrones onto observed CMDs agree with the
observed abundances. To answer such questions as ``Do GC giants with the reddest 
$(M_{F438W}-M_{F606W})_0$ colours have [C/Fe] $\sim +0.5$?", as implied by our
isochrones, it is important to check such predictions spectroscopically.  This
is necessary because the model $\teff$ and hence colour scales are subject to
many uncertainties (such as the treatment of convection and the atmospheric
boundary condition).  In fact, spectroscopic studies of member stars with the
reddest and bluest $(M_{F336W}-M_{F438W})_0$, $(M_{F438W}-M_{F606W})_0$, and
$(M_{F606W}-M_{F814W})_0$ colours are needed to validate the models.  These
samples should consist of isolated stars with the best possible photometry so
that the colour spreads at a given magnitude can be defined to very high
accuracy.  Ideally, for consistency reasons, the spectroscopic analyses should
employ the same MARCS model atmospheres and synthetic spectra that we have used, 
and they should adopt temperatures and gravities very smiilar to those given
by the stellar models.  As we have emphasized throughout this study, comparisons
of predicted and observed variations in colour at a given absolute magnitude
should be more trustworthy than fits to the colours of individual stars.

A related issue that warrants some thought is whether the inferred abundances
would be very different had we employed BCs based on 3D, instead of 1D, model 
atmospheres.  It is well known that the strengths of molecular features are
quite sensitive to 3D effects, mainly because of differences in the temperature
structures of the outer atmospheres (see, e.g., \citealt{cat06}, \citealt{hac11},
\citealt{ana19}).  Because the outer layers are cooler in 3D atmospheres of
metal-poor stars, observed line strengths can be reproduced on the assumption of
lower abundances of the metals (at fixed $\teff$, $\log\,g$, and [Fe/H]) than in
the case of 1D models.  For instance, Hayek et al.~(see their Figs.~18, 19) have
found from their study of 3D models of the atmospheres of giants with $\log\,g =
2.2$ that fits to CH, NH, and OH spectral features in the UV result in negative
abundance corrections for C, N, and O, respectively --- ranging up to as much as
several tenths of a dex, with larger corrections occurring at the lower [Fe/H]
values.  

However, the extent to which such findings apply to the chemical abundances
inferred here is not clear.   The use of 3D atmospheres as boundary conditions
for stellar interior models is bound to have some effect on the predicted
$\teff$ scale, which is not constrained by empirical determinations to better
than $\sim\pm 70$--100~K; see, e.g., the compilation of temperatures and their
uncertainties for the sample of dwarfs and giants given by \citealt{crm10}).
Moreover, \citet{ccc18} have shown that BCs for broad-band filters in the UV
are affected at the level of up to a few hundredths of a magnitude, though
the differences are much smaller for optical and IR passbands.  Still, this is
an important issue that should be investigated in due course.

The encouraging success that we have had in explaining the photometric
properties of stellar populations in GCs with different abundances of C, N, and
O gives us some optimism that our models will provide improved interpretations
of chromosome maps.  However, before applying our models to such maps, it will
be necessary to extend the MARCS spectra further into the UV so that BCs for
the $F275W$ filter can be generated (and subsequently tested), and to produce
model atmospheres, synthetic spectra, BCs, and stellar models for wider ranges
in the nitrogren abundance as well as for additional C:N:O ratios to sample the
observed C--N and O--N anticorrelations in more detail.  Once the additional
models are in hand, it should be possible to place reasonably tight constraints
on the absolute light element abundances of stars in the most populous stellar
populations and sub-populations that have been identified in GCs.

\section*{acknowledgements}
We thank Kjell Eriksson for investigating the interplay between model
atmospheres and synthetic spectra, Poul Erik Nissen and Anish Amarsi for valuable
comments and suggestions after reading the initial version of this paper,
and Karsten Brogaard, Pavel Denisenkov, John Norris, Peter Stetson, and David
Yong for helpful information on various aspects of this project and for
mentioning a few papers that we had overlooked.  LC is the recipient of the ARC
Future Fellowship FT160100402.

\section*{Data Availability}
Reference should be made to Paper I for information on how to obtain 
a selection of isochrones for [Fe/H] $= -0.5, -1.5$, and $-2.5$ and all of the
metal abundance mixtures that have been considered in this project, along with
the means to transpose them from the theoretical H-R diagram to various CMDs.
Grids of evolutionary tracks for finer spacings of [Fe/H], assuming two or more
helium abundances at each metallicity, are in the process of being computed;
they will be made available to interested users once a separate paper
describing these models has been submitted for publication.

\bsp   
\label{lastpage}

\end{document}